\documentclass[10pt,journal,compsoc]{IEEEtran}

\usepackage{amssymb}
\usepackage{graphicx}
\usepackage{amsmath}
\usepackage{psfrag}
\usepackage{multirow}
\usepackage{url}

\ifCLASSOPTIONcompsoc
  \usepackage[nocompress]{cite}
\else
  \usepackage{cite}
\fi
\ifCLASSINFOpdf
\else
\fi
\begin{document}
%
\title{Exact Inference Techniques for the Analysis of Bayesian Attack Graphs}
%
%
%
%

\author{Luis~Mu\~noz-Gonz\'{a}lez,
        Daniele~Sgandurra,
        Mart\'{i}n~Barr\`ere,
        and~Emil~C.~Lupu
\IEEEcompsocitemizethanks{\IEEEcompsocthanksitem All the authors are with the Department of Computing at Imperial College London, 180 Queen's Gate, SW7 2AZ, London, UK.\protect\\
E-mail: \{l.munoz, d.sgandurra, m.barrere, e.c.lupu\}@imperial.ac.uk}
}

\IEEEtitleabstractindextext{%
\begin{abstract}
Attack graphs are a powerful tool for security risk assessment by analysing network vulnerabilities and the paths attackers can use to compromise network resources. The uncertainty about the attacker's behaviour makes Bayesian networks suitable to model attack graphs to perform static and dynamic analysis. Previous approaches have focused on the formalization of attack graphs into a Bayesian model rather than proposing mechanisms for their analysis. In this paper we propose to use efficient algorithms to make exact inference in Bayesian attack graphs, enabling the static and dynamic network risk assessments. To support the validity of our approach we have performed an extensive experimental evaluation on synthetic Bayesian attack graphs with different topologies, showing the computational advantages in terms of time and memory use of the proposed techniques when compared to existing approaches.
\end{abstract}


\begin{IEEEkeywords}
Security risk assessment, attack graphs, Bayesian networks, dynamic analysis, probabilistic graphical models.
\end{IEEEkeywords}}

\maketitle

\IEEEdisplaynontitleabstractindextext

%
\IEEEpeerreviewmaketitle

\IEEEraisesectionheading{\section{Introduction}\label{sec:introduction}}



\IEEEPARstart{T}{he} estimated cyber-security market is expected to grow to \$101 billions in 2018 \cite{gartner}. Nevertheless, efforts to protect networks cannot cope with the sophistication of attackers, as shown by the history of data-breaches organizations have suffered, including in recent times \cite{beautiful}. However, it is not always possible to patch all existing vulnerabilities: some systems cannot be interrupted or a lack of manpower prevents from doing so. Therefore, one way to optimize the resources and effort required to protect a network is to firstly assess its risks, and then, prioritize the most critical threats. This requires estimating the risk exposure of vulnerable network nodes, given the threat likelihood and the severity of the impacts \cite{wheeler}, and using these values to select appropriate countermeasures. This process produces not only a better threat prioritization, but also an improved return-on-investment. However, this approach does not consider the dependencies between vulnerabilities, thus limiting its usefulness.

These shortcomings can be addressed with Attack Graphs (AGs) \cite{sheyner2002,jha,albanese}, which represent prior knowledge about vulnerabilities and network connectivity, enabling system administrators to reason about threats and their risks in a formal way. AGs permit a priori analysis of the possible avenues an attacker can exploit to compromise the system. Thus, they can be used to focus on the most-effective threats and produce a better countermeasures selection \cite{ingols}, which is also known as \textit{static analysis}.

On the other hand, proactive security hardening is not always the best strategy. As discussed in \cite{barth}, a reactive security strategy can be competitive when the defender does not overreact to the last attack but learns from past experience. In this sense, AGs can also be used to \textit{dynamically} profile the attacker's paths, to determine which nodes are more likely to be attacked in the next steps. They can also be used to evaluate the security risks for valuable network resources and to reason about nodes that may have been already compromised, when we observe evidence of an ongoing attack. Since organizations are often under attack, this dynamic analysis gives system administrators important insights in real-time on where they should spend their efforts and the most vulnerable targets.

Both \textit{static} and \textit{dynamic analysis} of AGs have inherent probabilistic characteristics given the uncertainty about the attackers' ability to exploit vulnerabilities. In this sense, Bayesian Networks (BNs) provide an appropriate framework to model AGs, since they depict causal relationships between random variables in a compact way. This approach has already been proposed in the literature: \cite{liu} present a Bayesian AG (BAG) to model attack paths in a network, using Variable Elimination (VE) as an algorithm for inference on the Bayesian model. \cite{frigault,wang2008} present mechanisms to calculate the conditional probability tables, which represent the combined effect of vulnerabilities to compromise a node. More recently, \cite{poolsappasit} present a BN framework to perform risk assessment and propose risk mitigation strategies in the context of AGs. 

However, none of the above propose appropriate and efficient algorithms for inference on their models, and computing unconditional probabilities in BNs is an NP-Hard problem. For example, using a brute force approach and computing the joint probability distribution for a BAG with $40$ nodes, using $8$ bytes to store each entry in the table, requires $2^{40 + 3}/1,024^3 = 8,192$ Gigabytes of memory. Therefore, the use of efficient inference techniques is important to reduce the time and computational resources required and improve the applicability of the approach. More concretely, in \cite{frigault,wang2008} no mechanism is proposed to calculate the unconditional probabilities of compromising each node. Forward-backward propagation is proposed in \cite{poolsappasit}. However, as shown in  \cite{murphy,rabiner}, this technique is applicable only when the corresponding graph is a chain\footnote{In this context, a chain is a graph where, given nodes $X_1,..., X_N$, there is only one edge from each $X_i$ to $X_{i+1}$ for $i=1,..., N-1$.}, which is not true for AGs in general. Finally, although the VE algorithm proposed in \cite{liu} is valid for inference in BAGs, its computational complexity limits its applicability to small graphs, especially in the case of the dynamic analysis where the time to respond to an attack is of essence. Furthermore, none of the previous papers reports an experimental evaluation of the time and memory requirements of the techniques proposed to assess their suitability for static and dynamic analysis of AGs. 

The main contributions of this paper are the following:
\begin{itemize}
\item We propose a revised BAG model for the static and dynamic analysis of AGs, which overcomes some limitations of previous models, such as the undesirable effects of adding a prior on the attacker capabilities. Furthermore, this model serves as a basis for further model extensions, including zero-day vulnerabilities, attacker's capabilities or dependencies between vulnerability types, among others. 
\item  Although exact inference in probabilistic graphical models is NP-Hard, we propose to use message passing algorithms such as Belief Propagation (BP) (for Attack Trees) and Junction Tree (JT) (for general AGs), to efficiently calculate the unconditional probabilities that the nodes have been compromised considering, if applicable, evidence of ongoing attacks.
\item  To assess the applicability of the algorithms proposed and to show the limitations of existing approaches, we provide a comprehensive experimental evaluation using synthetic AGs. Our results show that the JT algorithm can be applied to AGs of hundreds of nodes, corresponding to networks of thousands of nodes. As far as we know, this is the first experimental evaluation in the literature of AGs that analyses the time and memory requirements for static and dynamic inference in BAGs.
\item Our results also show the importance of cluster structures when modelling with AGs. We show that the JT algorithm in clustered networks scales linearly in the number of nodes for dynamic inference, which makes it suitable for use in practical settings such as corporate networks where hosts are grouped for management purposes. 
\end{itemize}


The rest of the paper is organised as follows. Section 2 reviews Attack Graph models. In Section 3 we present a BAG model that improves upon existing models in the literature. In Section 4 we introduce VE, BP, and JT as procedures to perform exact inference on BAGs. Experimental results for static and dynamic inference on BAGs are presented in Section 5 while in Section 6 we sketch possible extensions of our model. Section 7 concludes the paper and discusses further research directions. 

\section{Attack Graphs}
AGs are graphical models that represent the knowledge about vulnerabilities in a network, and their interactions, showing the different paths an attacker can follow to reach a given goal. Along each path, vulnerabilities are exploited in sequence, each successful exploit giving the attacker more privileges towards his goal. Two main types of AG are used in the literature: \textit{state-based representations} and \textit{logical AGs}.

In \emph{state-based} representations \cite{jha,phillips,sheyner2002} each node in the AG reprents the state of the whole network after a simple atomic attack, and contains a table with global variables defining that state. The number of states and variables combinatorially explodes when increasing the number of nodes \cite{ammann,jajodia,ou2006}, thus limiting the applicability of these representations to very small networks only. 
Moreover, state-based AGs can contain duplicate attack paths that differ only in the order of the attack steps. This additionally increases the complexity of the graph.


In contrast, \emph{logical AGs} are defined as bipartite graphs which represent dependencies between exploits and security conditions \cite{ammann}. These representations rely on a monotonicity principle: the attacker never relinquishes privileges once obtained. Although not always applicable this assumption is reasonable in most cases. Monotonicity allows to remove duplicated paths and results in a Directed Acyclic Graph (DAG), which grows polynomially with the number of vulnerabilities and the number of connected pairs of hosts \cite{albanese}. Formally, we can define an AG (in the logical representation) as a directed bipartite graph $G = (E \cup C, R_r \cup R_i)$, where the vertices $E$ and $C$ are the sets of exploits and security conditions, respectively, and the edges $R_r \subseteq C \times E$ and $R_i \subseteq E \times C$  are \textit{require} and \textit{imply} relations.

\begin{figure}
	\centering
	\includegraphics[width=4cm,height=4cm]{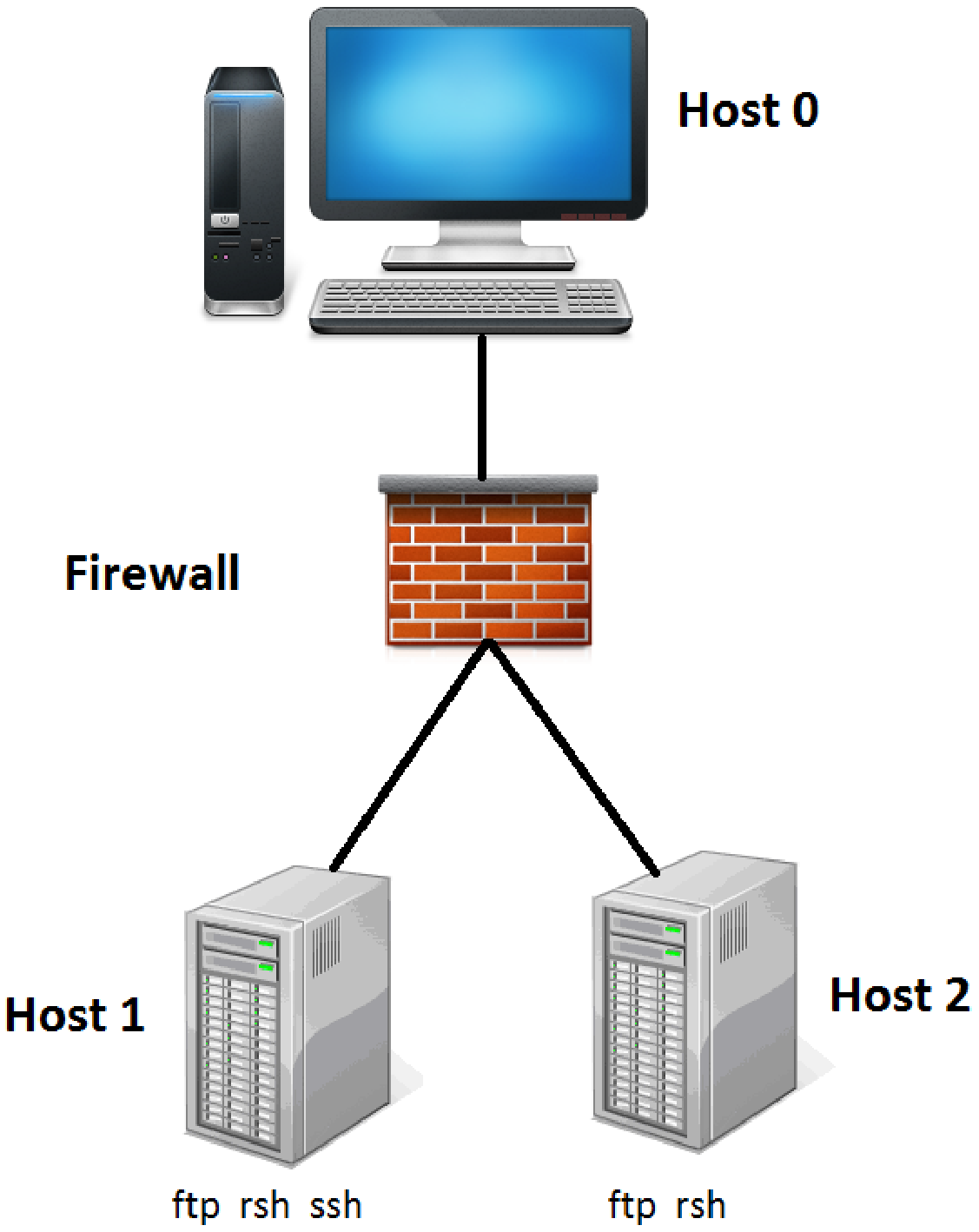}
	\includegraphics[width=4cm,height=6cm]{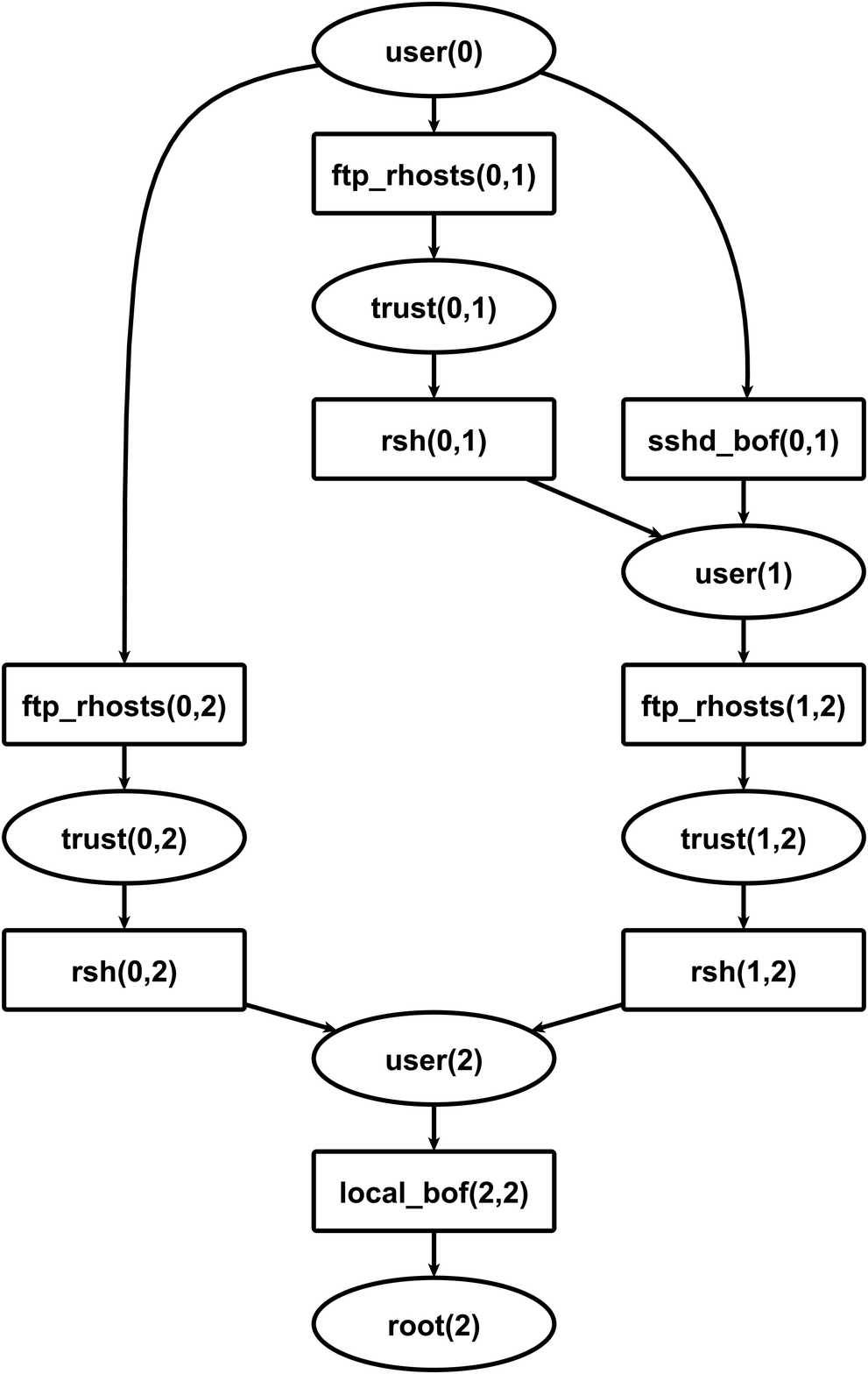}
	\caption{Simple example of a network configuration and the corresponding logical AG taken from \cite{frigault2,frigaultMsC}.}
	\label{fig1}	
\end{figure}

Figure \ref{fig1} shows a scenario from \cite{frigault2,frigaultMsC}, where \textit{Host 1} offers File Transfer Protocol (FTP), Secure Shell (SSH), and Remote Shell (RSH) access, whilst \textit{Host 2} offers FTP and RSH. The firewall allows FTP, SSH, and RSH traffic from external users (\textit{Host 0}) to both servers. The goal of the attacker is to gain \textit{root} privileges on \textit{Host 2}. In the AG, the conditions are represented as circles where the host involved is inside the parentheses, while vulnerabilities are depicted in rectangles, showing the source and destination host inside parentheses, i.e., \textit{(source, destination)}. In Figure \ref{fig1}, we observe that there are three possible paths for the attacker. The probabilities that an attacker can successfully exploit the vulnerabilities in the network given in \cite{frigaultMsC}, using the Base Score metric of the CVSS score, are: $0.8$ for \textit{ftp\_rhost}, $0.1$ for \textit{ssh\_bof}, $0.9$ for \textit{rsh}, and $0.1$ for \textit{local\_bof}. 





\section{Bayesian Attack Graphs}
As logical representations of AGs result in DAGs, Bayesian Networks (BNs) are suitable to model the AGs and perform static and dynamic analysis, to calculate the probability that an attacker can reach each state (condition) in the graph.

The use of BNs for AGs was first introduced in \cite{liu} for the dynamic analysis of AGs. The authors proposed to use the VE algorithm \cite{dechter} to calculate the probability that an attacker can reach a security state given prior knowledge of the state it had reached. They also propose to use the Most Probable Explanation (MPE) algorithm (relying on VE) to determine the nodes that have been possibly already compromised. However, VE can be computationally expensive compared to other inference algorithms such as Junction Tree (JT). Furthermore, the authors do not propose an elimination ordering algorithm before applying VE, which has significant impact on the algorithm's performance, as we show in Section 5. Moreover, finding the optimal elimination order turns out to be another NP-Hard problem \cite{arnborg}. \cite{liu} also lacks an experimental evaluation to assess the applicability of the algorithm in practice. Finally, we consider that the use of MPE is not appropriate in the context of AGs and can lead to misleading conclusions about the network state, as further discussed in Section 4.

\cite{wang2008} and \cite{frigault} show how to calculate the conditional probability tables in a Bayesian Attack Graphs (BAGs) as the combined effect of vulnerabilities in a network. In \cite{frigault2}, a Dynamic BN is proposed to also model temporal factors that affect the impact of the vulnerabilities, however they do not provide any mechanism for inference on their models. \cite{poolsappasit} proposes a risk management framework using BNs to assess at run-time the chances of a network compromise and select mitigation strategies. However, the authors propose to use forward-backward propagation for inference, which is not appropriate for general AGs, as this algorithm can only be applied to chains, not to general graphs \cite{murphy,rabiner}.

A BN is a directed graphical model where the nodes represent random variables and the directed edges represent dependencies between them, forming a DAG. Let ${\bf X} = \{ X_1, ..., X_n \}$ be a set of random variables (continuous or discrete). The joint probability distribution can be written as:
\begin{equation}
p({\bf X}) = \prod_{i = 1}^n p(X_i | {\bf pa}_i )
\label{eq1}
\end{equation} so that, under the BN representation, for each node $X_i$ there is a directed edge from each node in the set of parents nodes ${\bf pa}_i$ of $X_i$ pointing to $X_i$.  
For example, the joint probability distribution of the BAG in Figure \ref{fig2} can be written as:
\begin{equation}
\begin{split}
p(A,B,C,D,E,F,G) = & \ p(A) \ p(B|A) \ p(D|A) \ p(C|A,B)  \times \\
& p(E|C) \ p(F|D,E) \ p(G|F)
\end{split}
\label{eq2}
\end{equation}

\subsection{Model assumptions}
Following a treatment similar to others works in the literature on AGs, we will make here the following assumptions to build the BAG from the AG logical representation:
1) We consider that successfully exploiting a vulnerability in a given context (e.g., on a host) does not change the probability of exploiting the same or similar vulnerabilities in a different context (on another host). 
2) The probability of successfully exploiting a vulnerability remains constant in time e.g., the attacker does not improve his success during the attack. Although in \cite{frigault2} a dynamic network is proposed to model such aspects, the changes in probabilities are slow enough (i.e., days or weeks) to be considered constant. It is then better to recompute the model, than to increase the complexity of the model to deal with such dynamic aspects.
3) The attacker's capabilities are not considered or, at least, all the potential attackers are supposed to have the same skills and attack preferences.
4) We do not consider zero-day vulnerabilities, social engineering attacks and insider attacks.
5) We assume that the Intrusion Detection System (IDS) may not detect all the events of interest and that it does not trigger false alarms (or false alarms have been discarded following investigation).

Although these assumptions may seem restrictive, they are common in the literature on AGs. Furthermore, the efficient probabilistic inference mechanisms we propose can also be applied to other more flexible BAG models as discussed in Section 6. Under these assumptions, the nodes in the BAG represent the different security states that an attacker can reach. We model the behaviour of these states as Bernoulli random variables, so the probability of a node $X_i$ to be compromised is\footnote{To simplify the mathematical notation, we will refer to the unconditional probability of a node to be compromised as $\text{Pr}(X_i)$.} $\text{Pr}(X_i = T) = p$, and, consequently, the probability of a node not to be compromised is  $\text{Pr}(X_i = F) = 1 - p$, with $p \in [0,1]$.

\begin{figure}
	\centering
	\includegraphics[width=5cm,height=4cm]{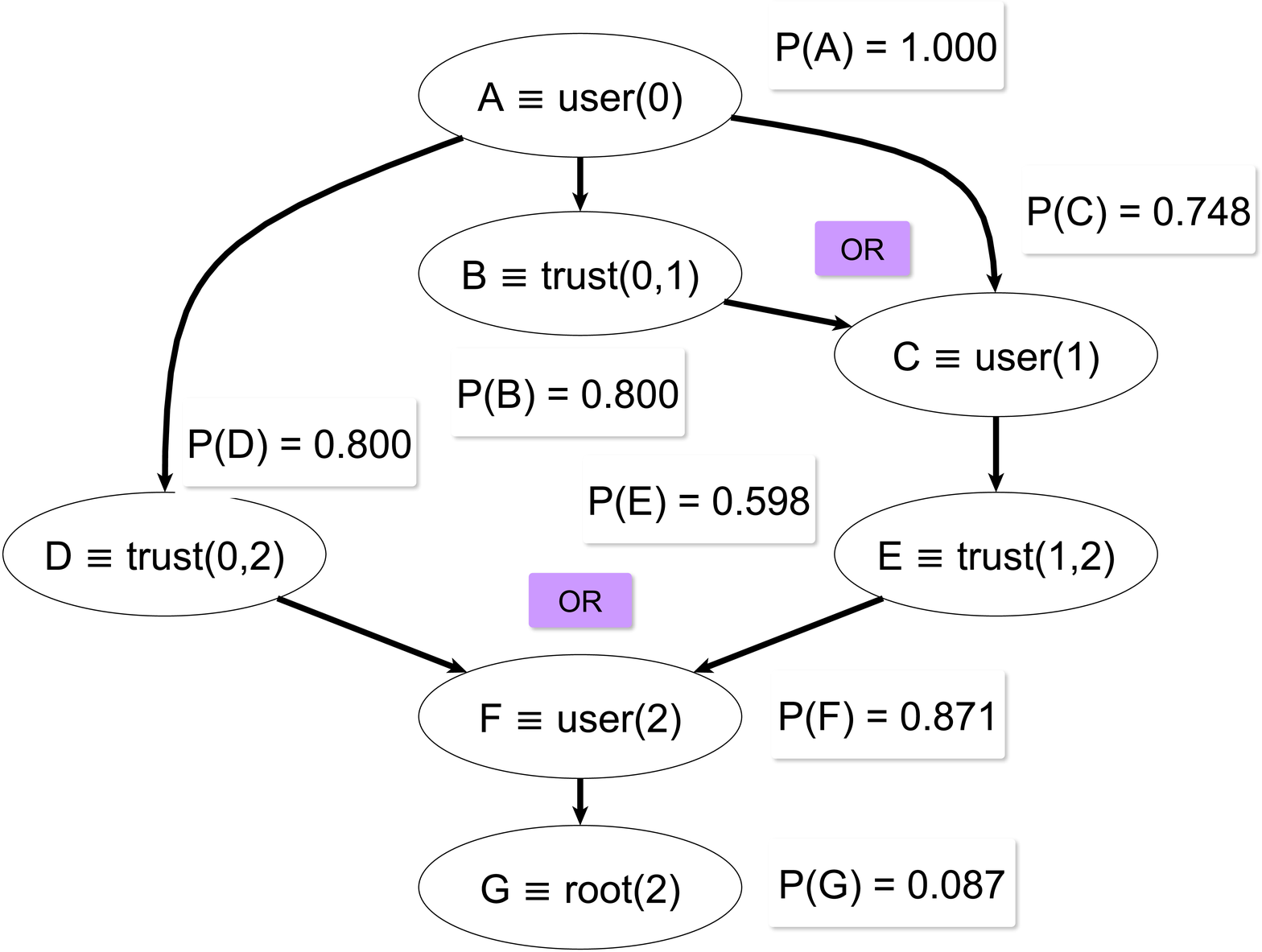}
	\caption{BAG representation for the AG in Figure \ref{fig1} with the unconditional probabilities calculated for each node when there are no attacks.}
	\label{fig2}	
\end{figure}

In Figure \ref{fig2} we show the BAG generated from the AG shown in Figure \ref{fig1} along with the probabilities for each node to be compromised by an attacker. The initial node, $A \equiv \emph{user}(0)$, shows that the attacker has user privileges on his own machine with probability $1$. 

\subsection{Conditional Probability Distributions}
Under the BN representation, the information available at each node $X_i$ is the conditional probability distribution of $X_i$ to be compromised given its parent nodes, i.e., $p(X_i | {\bf pa}_i)$. From a security view point, these conditional probabilities represent the probabilities of an attacker to reach security state $X_i$ given the observations of the set of preconditions ${\bf pa}_i$. These allow the attacker to compromise $X_i$ by exploiting the vulnerabilities ${\bf e}_i$, which link ${\bf pa}_i$ with $X_i$ in the original bipartite AG. We consider that the probabilities of successfully exploiting vulnerabilities are parameters of the model (instead of random variables), which allows to calculate the conditional probability tables that define $p(X_i | {\bf pa}_i)$.

The scores provided by the Common Vulnerability Scoring System (CVSS) \cite{cvss} can be used to estimate $p_{v_j}$, the probability of an attacker successfully exploiting a vulnerability $v_j$. Although CVSS scores estimate the impact of a vulnerability rather than its probability of being successfully exploited, in the absence of better indicators, CVSS scores or some of their submetrics are often used in the literature. Whilst \cite{frigault2,houmb} use the entire CVSS score, in our opinion, the exploitability submetric is more appropriate since it tries to measure the difficulty of exploiting a vulnerability. This is also proposed in \cite{poolsappasit}. For the AG in Figure \ref{fig1} we have used the probabilities given by \cite{frigaultMsC}, which only consider the Base Score metric of the CVSS score.

To calculate the conditional probability distributions $p(X_i | {\bf pa}_i)$ 
we consider two possible cases \cite{poolsappasit}:
A logical \textit{AND} where all the preconditions should be met to compromise node $X_i$. This can be expressed as:
\begin{equation}
p(X_i | {\bf pa}_i) = \begin{cases} 0, & \exists X_j \in {\bf pa}_i | X_j = F \\ \prod_{j: X_j} p_{v_j}, & \text{otherwise} \end{cases}
\label{eq4}
\end{equation}
A logical \textit{OR} where only one of the preconditions in ${\bf pa}_i$ needs to be satisfied to compromise $X_i$. This can be calculated using the noisy-OR formulation \cite{koller}, so that:
\begin{equation}
p(X_i | {\bf pa}_i) = \begin{cases} 0, & \forall X_j \in {\bf pa}_i | X_j = F \\ 1 - \prod_{j: X_j} (1 - p_{v_j}), & \text{otherwise} \end{cases}
\label{eq5}
\end{equation}

The Supplementary Material shows how to compute the conditional probability tables for the \text{AND} and \text{OR} cases through an example.

\begin{figure}
	\centering
	\includegraphics[width=4.7cm,height=3.5cm]{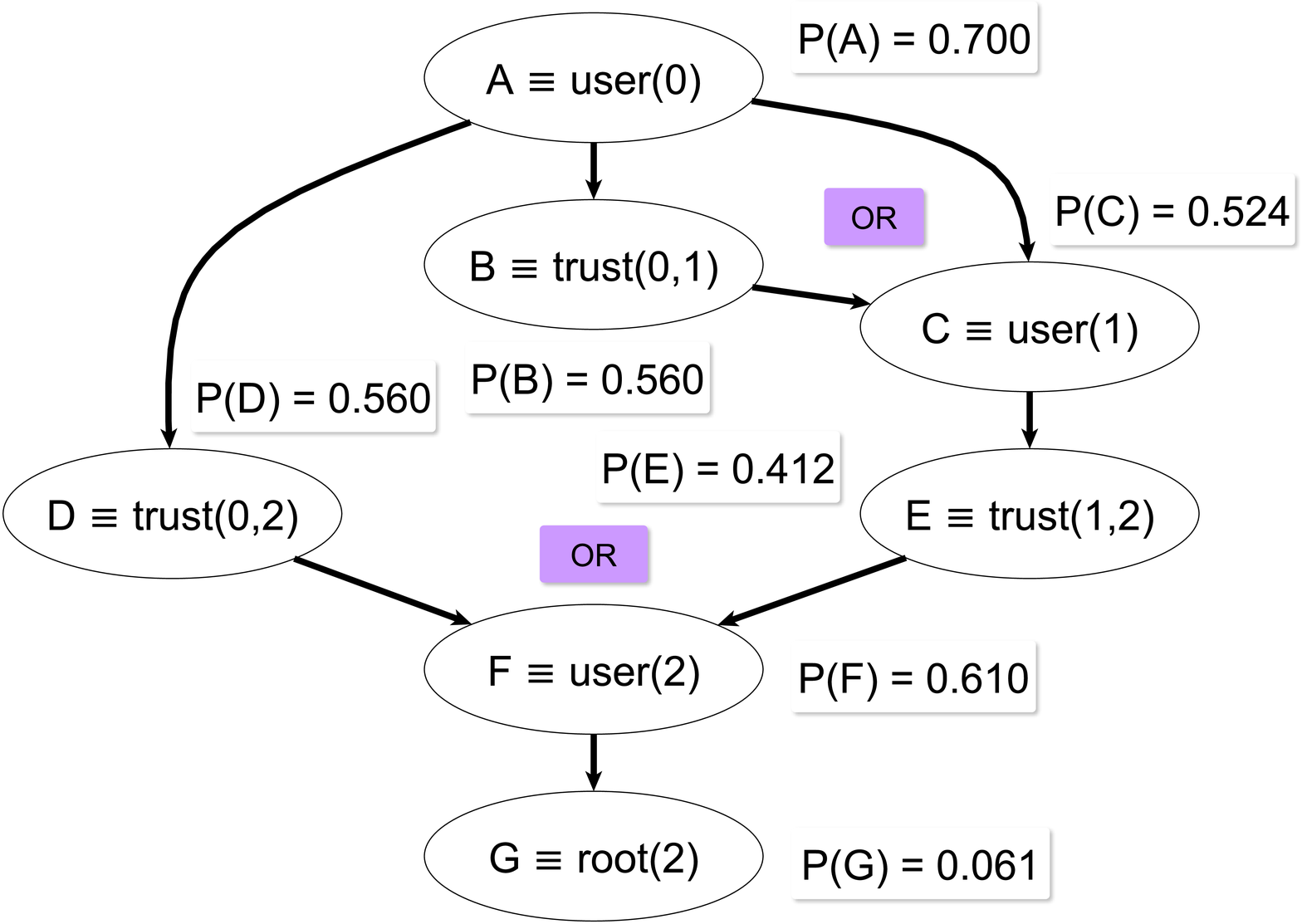} \\
	(a) \vspace{0.5cm} \\
	\includegraphics[width=4.7cm,height=3.5cm]{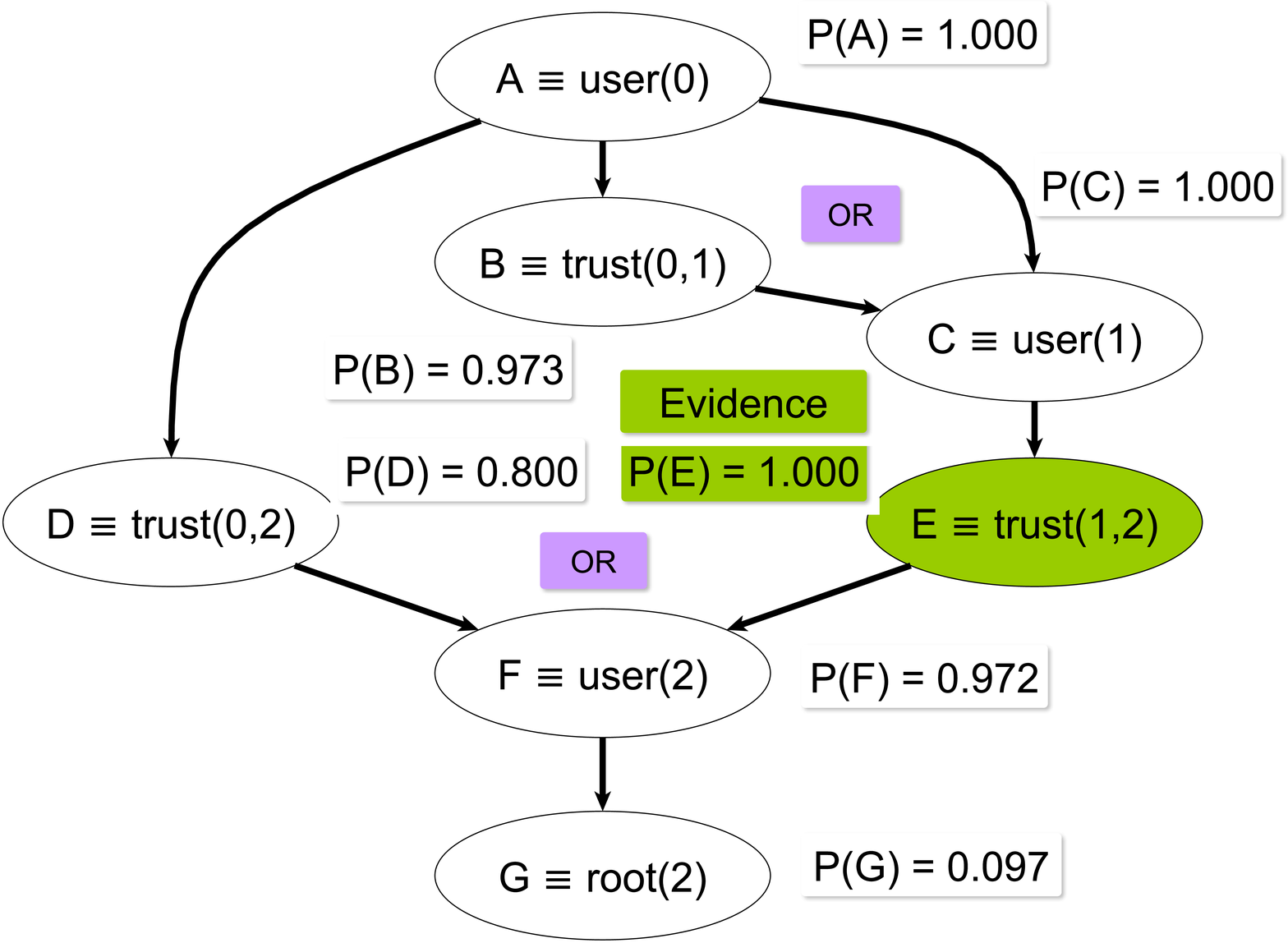}  \\
	(b)  \\
	\caption{(a) Unconditional probabilities for the BAG in Figure \ref{fig2} when the prior belief on the initial state of the attacker (node A) is set to $0.7$. (b) Unconditional probabilities for the previous BAG when evidence that the attacker has compromised node $E$ is observed. The result is the same as when observing the same evidence for the BAG in Figure \ref{fig2}.}
	\label{fig3}	
\end{figure}

\subsection{Effect of the prior probability on the initial state}
The effect of the initial node of the BAG, which represents the initial state of the attacker (for example, node $A$ in Figure \ref{fig2}) requires further consideration. It has not been discussed in the literature and can have significant implications for the analysis of AGs, as we will show. In our opinion, this node does not represent a random variable, as considered in other approaches, since it only represents that the attacker has full rights on his own machine. This is equivalent to consider that $\text{Pr}(X_0) = 1$. Other studies, e.g. \cite{poolsappasit}, propose to use this node to reflect some subjective prior knowledge of the attacker's capabilities or the attacker intention and let the administrator set the value of $\text{Pr}(X_0)$. This has two main shortcomings: firstly, modelling the attacker's capabilities only describes a subjective average behaviour of different kinds of attackers and, secondly, the effect of this prior can lead to misleading conclusions in the dynamic analysis of the AG, especially when reasoning about the nodes an attacker may have already compromised.

To illustrate this, consider the example in Figure \ref{fig3}.(a), where we have the same BAG as in Figure \ref{fig2}, but with the prior belief on the initial state to the attacker set to $\text{Pr}(A) = 0.7$, instead of $1$. As expected, the unconditional probabilities of the other nodes decrease with respect to the probabilities calculated in Figure \ref{fig2}. At some stage, forensic evidence of an attack may let us conclude that node $E$ has been compromised as shown in Figure \ref{fig3}.(b); thus $\text{Pr}(E) = 1$. 
Note that this result is the same regardless of the value of $\text{Pr}(A)$. In the case where $\text{Pr}(A) = 1$, we see that all nodes have increased their probabilities except $D$. This indicates that nodes $C$ and $B$, as parents of $E$, may have been compromised. In contrast, if $\text{Pr}(A) = 0.7$ and $E$ has been compromised, the probabilities of \textbf{all} the nodes have increased. This is misleading, as there is no additional evidence that the attacker followed the path from $A$ to $D$ or that the attacker's ability to compromise $D$ has increased. This effect can hinder reasoning about the attack paths that an attacker could follow or nodes that may have been already compromised. 

\begin{figure}
	\centering
	\includegraphics[width=4.7cm,height=3.7cm]{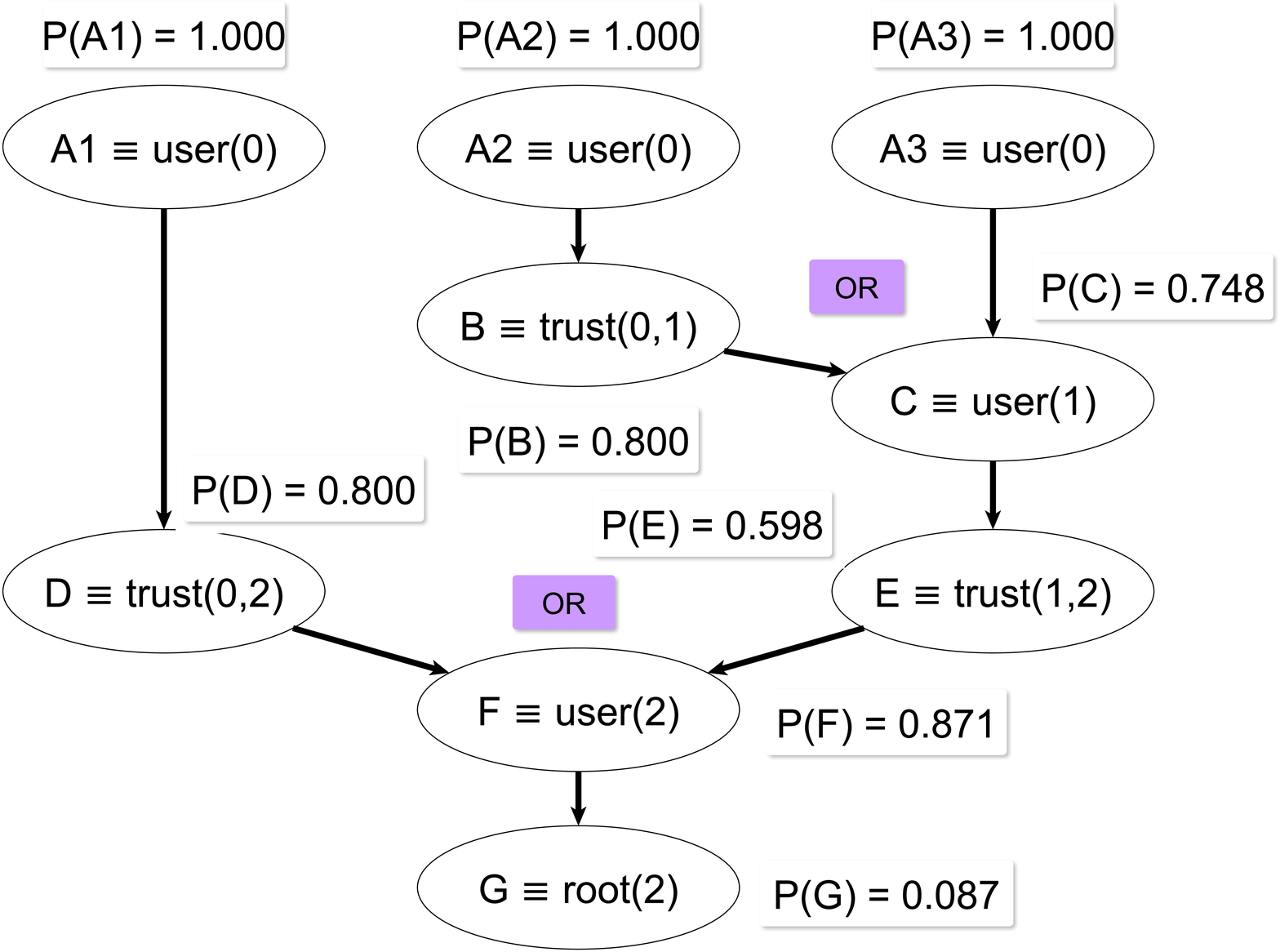} \\
	(a) \vspace{0.4cm} \\
	\includegraphics[width=4.7cm,height=3.7cm]{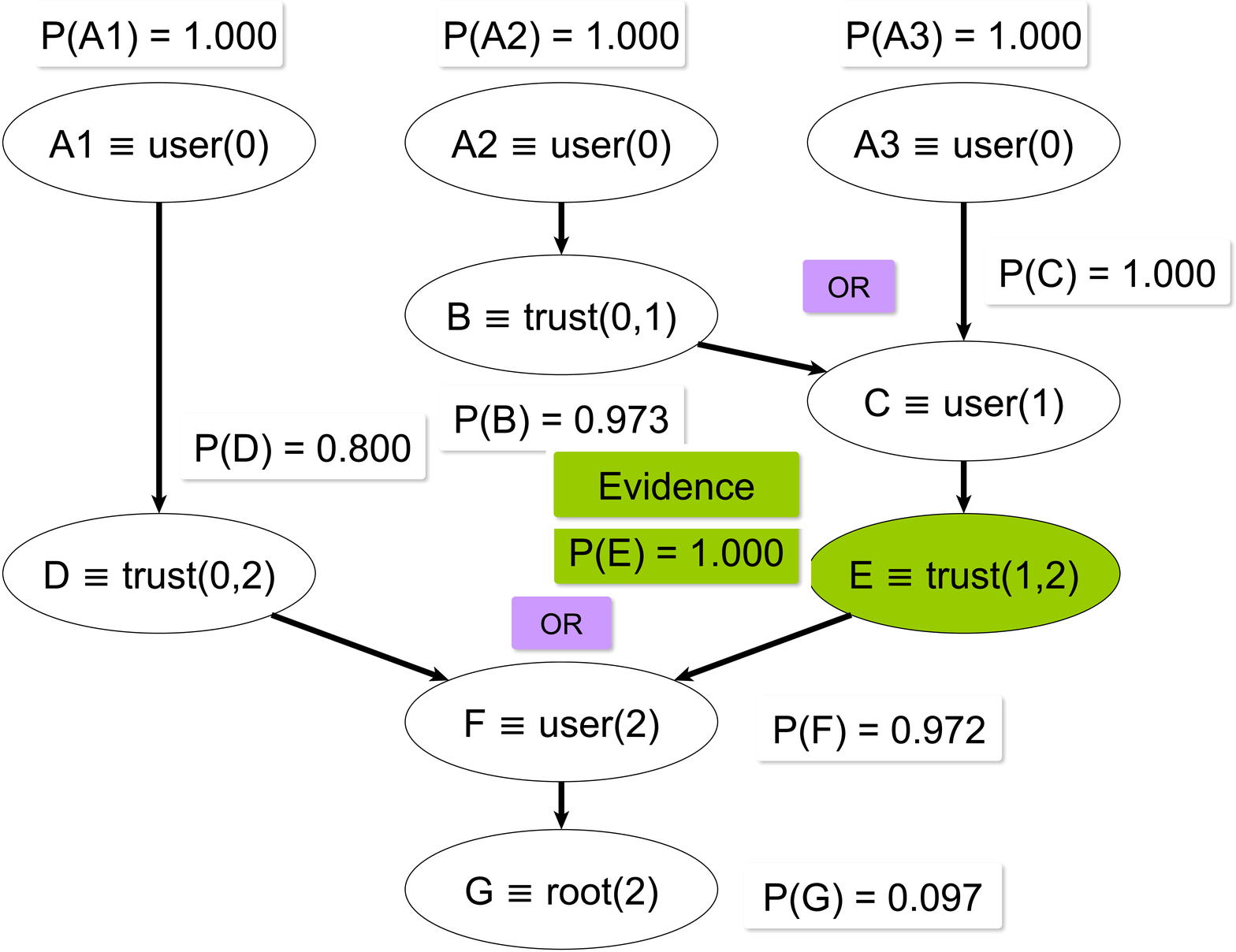}  \\
	(b)  \\
	\caption{(a) Unconditional probabilities for the BAG in Figure \ref{fig2} splitting the node of the initial state of the attacker in 3 nodes, one for each initial attack path. (b) Unconditional probabilities for the previous BAG when evidence that the attacker has compromised node $E$ is observed. The result is the same than in Figure \ref{fig3}.(b).}
	\label{fig4}	
\end{figure}

Finally, to calculate the unconditional probabilities of the BAG more efficiently, we propose to use one initial node (the initial state of the attacker) for each initial attack path. This does not affect the values of the unconditional probabilities for the other nodes (with and without evidence of possible compromise), but allows to break some loops in the graph, which reduces the complexity of the inference algorithms described in the following sections. For example, the BAG in Figure \ref{fig2} becomes the tree shown in Figure \ref{fig4}. This makes it suitable to use Belief Propagation (BP) to calculate the unconditional probabilities of all the nodes efficiently. Note also that by splitting the initial state of the attacker in different nodes, we consider the initial attack paths independently and can thus track their change in probabilities when considering new evidence of compromise. This recommendation applies even when the prior belief on the attacker's state is not set to $1$ and represented as a random variable and avoids the misleading intuitions we can get when modelling the initial state as in Figure \ref{fig3}.(a).

\subsection{Applying BAGs for Security Risk Assessment}

The applicability of the proposed BAG model for security risk assessment can be categorized in: \textit{static risk analysis}, \textit{dynamic risk analysis}, and \textit{dynamic risk mitigation}.

For \textit{static risk analysis} we consider the security posture at rest. From the network topology, network reachability and the results of a vulnerability analysis we can build the BAG model, determine the values of the successful exploitation of vulnerabilities e.g., using the Base Score metric of the CVSS score, and build the conditional probability tables. Then, using inference techniques, such as the JT algorithm, we can compute the unconditional probabilities of all the nodes in the BAG. These probabilities serve as risk estimates that can be used to detect weak areas in the network and serve as an input for network hardening or static risk mitigation techniques.

For \textit{dynamic risk analysis}, the BAG model is recomputed at run-time, taking into account indications that some of the network components may have been compromised, e.g., from a Security and Incident Event Management System (SIEM) or IDS. Then, the nodes where we observe evidence of compromise are set to 1 (or to a value corresponding to the evidence observed) and the posterior probabilities of the rest of the nodes in the BAG given the evidence are computed. In this context, our model enables administrators to dynamically profile the possible attack paths that the attacker is following and to determine the nodes that are more likely to be compromised in the next steps. 

The computed posterior probabilities also provide the administrator with run-time risk estimates that can be used to select countermeasures and \textit{dynamic risk mitigation} strategies. This involves planning the most efficient strategies to reduce the risk taking into account the available security measures that can be applied and the cost of applying them. Once countermeasures have been selected and applied the conditional probabilities can be updated in the BAG and the posterior probabilities recomputed. We do not consider in this paper methods for the selection of countermeasures but our BAG model can be combined with that proposed in \cite{poolsappasit}, which models risk mitigation as a discrete reasoning problem solved using a genetic algorithm. 

\section{Exact Inference in BAGs}
For the analysis of AGs, we are interested in calculating the unconditional probability distributions $p(X_i)$, rather than $p(X_i | {\bf pa}_i)$, to determine the probability that an attacker can reach a given security condition, and thus, the risk. Using Bayes rule it is possible to calculate $p(X_i)$ from the conditional probability distributions:
\begin{equation}
p(X_i) = \sum_{{\bf X} - X_i} p({\bf X})  = \sum_{{\bf X} - X_i} \prod_{j=1}^n p(X_j | {\bf pa}_j )
\label{eq6}
\end{equation} where ${\bf X} - X_i$ indicates that we sum over all the set of random variables ${\bf X}$ except $X_i$. 

However, the exact calculation of (\ref{eq6}) is an NP-Hard problem \cite{cooper,koller}. In our case, as each node corresponds to a Bernoulli random variable, the memory required to store the joint probability distribution $p({\bf X})$ grows as $2^n$.  Thus, applying brute force for inference in probabilistic graphical models is not a reasonable approach in terms of computational time and memory, even for small graphs, and the use of efficient algorithms is a strong requisite.


As discussed in previous section, this issue has not been covered adequately in the literature on BAGs. \cite{poolsappasit} proposes to use forward-backward propagation, but this procedure is only valid for chains \cite{murphy,rabiner} and cannot be applied to general AGs. The Variable Elimination (VE) algorithm proposed in \cite{liu}, is an efficient technique to calculate the unconditional probabilities, but the authors do not propose a heuristic to find a reasonable elimination ordering, which impacts the performance of the algorithm significantly. VE is also used in \cite{liu} for Maximum A Posteriori (MAP) estimation to provide the MPE. However, MAP estimations can lead to misleading conclusions. For the example in Figure \ref{fig2}, the result of the MPE queries when there is no evidence of attacks is that all the nodes except $G$ are in the True state (and then $G$ is in the False state). The attacker would have therefore already compromised all the nodes except $G$, which makes no sense. In this case, we can easily show that MPE is impractical to assess the security risk in AGs.


In the next section, we first review VE, the algorithm that was used in \cite{liu}. Then, we describe the Belief Propagation (BP) and Junction Tree (JT) algorithms, which we propose for static and dynamic analysis of BAGs. These algorithms use a message passing approach to calculate the unconditional probabilities on BNs and their average computational complexity is significantly lower than VE.

\subsection{Variable Elimination}
VE or Bucket Elimination is a heuristic first introduced in \cite{zhangPoole94} and revisited in \cite{dechter} to efficiently compute the unconditional probabilities in BNs and Markov Random Fields (MRFs). In essence, to address the exponential blow-up when computing marginal probabilities, it identifies factors in the joint distribution that depend on a small number of variables, computes them once and caches the results to avoid generating them exponentially many times \cite{koller}.

For example, consider the probability $p(G)$ that an attacker can obtain root privileges on \textit{Host 2}, in the AG shown in Figure \ref{fig2}. We can then write $p(G)$ as:
\begin{equation}
\begin{split}
p(G) = & \sum_A \sum_B \sum_C \sum_D \sum_E \sum_F p(A) \ p(B|A) \ p(C|A,B) \ \times \\
& \times \ p(D|A) \ p(E|C) \ p(F|D,E) \ p(G|F)
\end{split}
\label{ve1}
\end{equation} As discussed before, computing the joint distribution scales in time and memory as ${\cal O}(2^n)$, with $n=7$ in this case. In contrast to this brute force approach, VE groups factors that involve the same variables and marginalizes (sums over) those variables. Then, we can re-write $p(G)$ as:
\begin{equation}
\begin{split}
p(G) = & \sum_F p(G|F) \sum_E \sum_D p(F|D,E) \sum_C p(E|C) \ \times \\
& \times \ \sum_B \sum_A p(A) \ p(B|A) \ p(C|A,B) \ p(D|A)
\end{split}
\label{ve2}
\end{equation} Evaluating this expression from right to left we can recursively eliminate all the variables in the BN except $G$. In this case, we follow the elimination ordering $\boldsymbol{\Omega} = \{A,B,C,D,E,F\}$. The steps of elimination using the VE algorithm are shown in Table \ref{tab2}: At each step we create a new factor $\phi_i$ by multiplying all the factors that involve the variable we want to eliminate, and then marginalizing the corresponding variable from the factor $\phi_i$.

\begin{table}[ht]
\caption{Steps of VE algorithm for the BAG in Figure \ref{fig2} to calculate $p(G)$ using the elimination order ${\boldsymbol \Omega} = \{A, B, C, D, E, F\}$.} 
\label{tab2}
\begin{center}
\begin{small}
\begin{tabular}{|c|c|}
\hline
Var. & Factors \\
\hline
\hline
\multirow{ 2}{*}{$A$:} & $\phi_1(A,B,C,D) = p(A) p(B|A) p(C|A,B) p(D|A)$ \\ 
& $\tau_1(B,C,D) = \sum_A \phi_1(A,B,C,D)$ \\ \hline
\multirow{ 2}{*}{$B$:} & $\phi_2(B,C,D) = \tau_1(B,C,D)$ \\
& $\tau_2(C,D) = \sum_B \tau_1(B,C,D)$ \\ \hline
\multirow{ 2}{*}{$C$:} & $\phi_3(C,D,E) = \tau_2(C,D) p(E|C)$ \\ 
& $\tau_3(D,E) = \sum_C \phi_3(C,D,E)$ \\ \hline
\multirow{ 2}{*}{$D$:} & $\phi_4(D,E,F) = \tau_3(D,E) p(F|D,E)$ \\
& $\tau_4(E,F) = \sum_D \phi_4(D,E,F)$ \\ \hline
\multirow{ 2}{*}{$E$:} & $\phi_5(E,F) = \tau_4(E,F)$ \\
& $\tau_5(F) = \sum_E \phi_5(E,F)$ \\ \hline
\multirow{ 2}{*}{$F$:} & $\phi_6(F,G) = \tau_5(F) p(G|F)$ \\
& ${\bf p(G)} = \tau_6(G) = \sum_F \phi_6(F,G)$ \\
\hline
\end{tabular}
\end{small}
\end{center}
\end{table}

The same principle applies when we observe evidence of compromise on some of the nodes and compute the posterior probabilities of the nodes given the evidence. In this case, as described in \cite{koller,murphy}, we first compute the joint probability distribution of the query variable and the evidence, and then, divide by the marginal probability of the observed evidence. For example, if we observe that an attacker has compromised node $C$ in Figure \ref{fig2} (the attacker has user privileges on \emph{Host 1}), the posterior probability of $G$ given the evidence is calculated as:
\begin{equation}
p(G | C = 1) = \frac{p(G, C = 1)}{p(C=1)}
\label{ve5}
\end{equation}

The computational cost of the algorithm is exponential in the scope of the factor with the maximum number of variables created during the elimination. In the example shown in Table \ref{tab2}, this is $\phi_1$, whose scope is $A, B, C, D$. Its computation requires a table with $2^4 = 16$ entries, whereas the computation of the expression in (\ref{ve1}) requires a table with $2^7 = 128$ entries. So, even in this simple example we can notice significant savings.

The variables' elimination order has a significant impact on the size of the intermediate factors created, and thus, the computational complexity of the algorithm \cite{fishelson,koller}. For instance, if in the previous example we use the elimination ordering: ${\boldsymbol \Omega'} = \{B, A, C, D, E, F\}$, the factor created to first eliminate $B$ is $\phi_1'(A,B,C) = p(A) \ p(B|A) \ p(C|A,B)$, so that $\tau_1'(A,C) = \sum_B \phi_1'(A,B,C)$. Then, $\phi_2'(A,C,D) = p(A) \ p(C|A,B) \ \tau_1'(A,C)$ and $\tau_2'(C,D) = \sum_A \phi_2'(A,C,D)$. Following then the same elimination steps as in Table \ref{tab2} we obtain an ordering where the maximum scope of the biggest factor is reduced from 4 to 3. 

Although finding the elimination order that minimizes the scope of the biggest factor is also NP-Hard \cite{koller}, several greedy heuristics \cite{fishelson,huang,kjaerulff,koller} provide good elimination orders at reduced computational cost. These rely on the concept of an \textit{induced graph}: the graph obtained when eliminating a variable from the original one. They seek orderings that induce small graphs, thus eliminating variables so that the scope of the intermediate factors $\phi_i$ and $\tau_i$ remains as small as possible. Criteria commonly used in these heuristics include:\cite{fishelson,koller}:
1) \textit{Min-neighbours}: At each step removing the node with the fewest neighbours in the current graph;
2) \textit{Min-fill}: At each step removing the node whose removal requires adding the fewest edges in the induced graph;
3) \textit{Min-weight}: At each step removing the node with the minimum product of weights of its neighbours, where the weights are the number of elements in the scope of the conditional probability associated with the node (for example, in the BAG in Figure \ref{fig2}, the weight for node $E$ would be $3 \times 3 = 9$, since its neighbours $C$ and $F$ have $3$ variables in their scope);
4) \textit{weighted-min-fill}: At each step removing the node with the smallest sum of weights of the edges that need to be added to the graph due to its elimination (the weight of an edge is the product of the weights of the nodes associated to that edge).

The use of VE for BAGs was previously proposed in \cite{liu} using the algorithm from \cite{dechter}, where no elimination ordering is proposed. In the experiments described in Section 5, we will show the impact of the elimination ordering in terms of the time and memory required to compute the unconditional probabilities. 

Finally, as discussed in \cite{murphy}, the main disadvantage of VE, in addition to the exponential scalability with the scope of the biggest factor, is its inefficiency when we compute multiple queries, e.g., when we calculate the unconditional probabilities of all the graph nodes. In VE we need to compute the elimination ordering and the factors each time we make a query, whereas other algorithms like BP or JT cache information (messages) that can be re-used to efficiently compute the marginal probabilities as we explain below.


\subsection{Belief Propagation}
Like VE, BP allows to efficiently compute the unconditional probabilities in BNs and MRFs when the graph is a tree or a polytree. Although this is not the general structure of AGs, this technique can be applied to Attack Trees (ATs) \cite{schneier}. BP is also referred to as the \textit{sum-product} algorithm and is based on probabilistic message passing. This also is the basis of the JT algorithm which can be applied to any kind of BN and that we describe in Section 4.3. The first version of BP was proposed in \cite{pearl82} and, then, extended in \cite{kim} for the case of polytrees, although its complete formulation is only introduced in \cite{pearl}. 

To describe the algorithm we follow an approach similar to \cite{bishop}, based on factor graph representations \cite{frey,kschischang}. As explained before, BNs (and MRFs) express the joint probability distribution of several random variables as the product of factors over subsets of those variables. Factor graphs make this decomposition explicit by introducing additional nodes for the factors themselves in addition to those representing random variables. This results in a bipartite graph. 

\begin{figure}
	\centering
	\includegraphics[width=3.5cm,height=4.0cm]{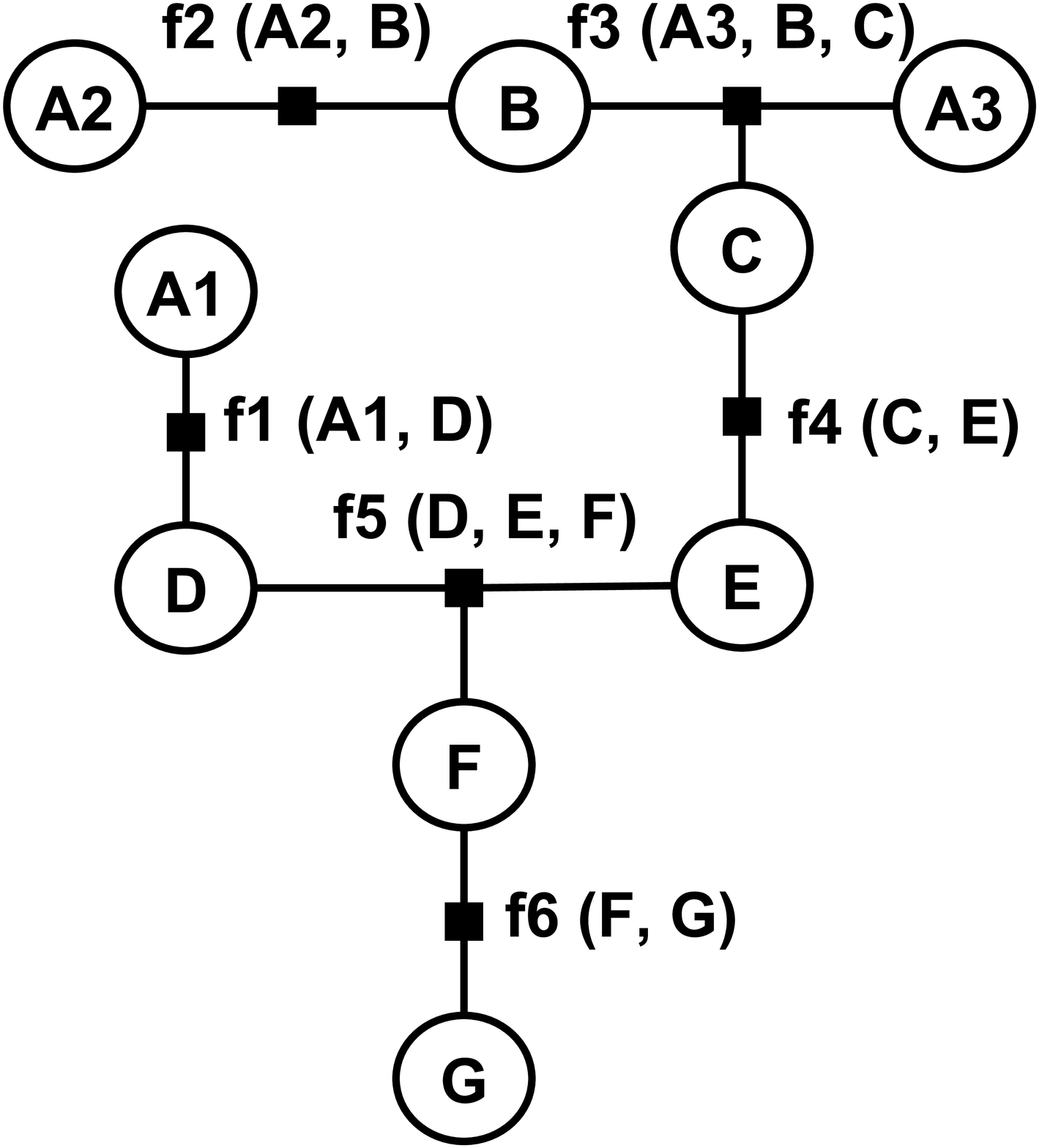}
	\caption{Factor graph for the BAG in Figure \ref{fig4}.}
	\label{fig5}	
\end{figure}

For example, the joint probability of the BAG shown in Figure \ref{fig4} can be expressed as:
\begin{equation}
p(A_1,A_2,A_3,B,C,D,E,F,G) = \prod_{i=1}^6 f_i ({\bf X}_i)
\label{bp1}
\end{equation} where the factors $f_i ({\bf X}_i)$ are:
\begin{equation}
\begin{split}
f_1(A_1, D) & = p(A_1) \ p(D| A_1) \\
f_2(A_2, B) & = p(A_2) \ p(B| A_2) \\
f_3(A_3, B, C) & = p(A_3) \ p(C|A_3, B) \\
f_4(C, E) & = p(E|C) \\
f_5(D, E, F) & = p(F| D, E) \\
f_6(F, G) & = p(G| F) \\
\end{split}
\label{bp2}
\end{equation} The factor graph is shown in Figure \ref{fig5}, though note that several factor graphs can exist for a given BN or MRF.

BP works by passing real valued functions called messages amongst neighbouring nodes in the tree or polytree network. Since the factor graph induces a bipartite graph, we can distinguish between two types of messages: from variable to factor and from factor to variable. Messages from a variable $X_i$ to a factor $f_j$ (in the neighbourhood of $X_i$) are given by:
\begin{equation}
\mu_{X_i, f_j} (X_i) = \prod_{f_k \in \{{\bf F}_i - f_j\}} \mu_{f_k, X_i} (X_i)
\label{bp3}
\end{equation} where $\mu_{f_k, X_i} (X_i)$ are the messages from the factor nodes in the neighbourhood of $X_i$ except $f_j$. Similarly, messages from a factor $f_i$ to a variable $X_j$ in the neighbourhood of $f_i$ are calculated as:
\begin{equation}
\mu_{f_i, X_j} (X_j) = \sum_{X_k \in {\bf X}_s} f_i(X_j,{\bf X}_s) \prod_{X_k \in {\bf X}_s} \mu_{X_k, f_j} (X_k)
\label{bp4}
\end{equation} where ${\bf X}_s$ is the set of variable nodes in the neighbourhood of $f_i$ except $X_j$.

When a variable $X_i$ is a leaf node, the corresponding messages to the factors in its neighbourhood are equal to one, i.e. $\mu_{X_i, f_j} (X_i) = 1$. On the contrary, if a factor $f_i$ is a leaf node, messages to a variable node in its neighbourhood are $\mu_{f_i, X_j} (X_j) = \sum_{X_k \in {\bf X}_s} f_i(X_j,{\bf X}_s)$.

BP computes messages from all the node variables to their corresponding factors and vice versa, starting from the leaf nodes and propagating the messages across the tree or polytree graph according to the following rule: A node $N$ (variable or factor) cannot send messages to another node $M$ until it receives all the messages from its neighbours except $M$. An example of the message passing process for the factor graph in Figure \ref{fig4} is shown in the Supplementary Material.

Once all messages are computed, the unconditional probability for a node $X_i$ can be calculated as:
\begin{equation}
p(X_i) = \prod_{f_j \in {\bf F}_i} \mu_{f_j,X_i} (X_i)
\label{bp5}
\end{equation} where ${\bf F}_i$ are the factor nodes in the neighbourhood of $X_i$. In MRFs, the same principle applies, but the product of the messages results in an unnormalized $p(X_i)$, though calculating the normalizing constant is straightforward. 

In contrast to VE, where we need to run the whole algorithm for each marginal probability we compute, BP can efficiently calculate all the marginal probabilities by computing all the messages once and storing them. So when we observe evidence of compromise in some nodes, only the factors that depend on the values that have changed need to be recomputed. For example, if the attacker has compromised node $D$, we take into account only the values of the function for which $D = T$, when computing messages involving factor $f_1$, so that
\begin{equation}
f_1(A_1, D = T) = p(A_1) \ p(D = T | A_1)
\label{bp7}
\end{equation} This requires us to consider only the elements of the conditional probability table for $p(D | A_1)$ where $D = T$. On the other hand, when computing the messages from factor to variable nodes involving observed variables, i.e., variables for which we observe new evidence, we do not need to sum over these variables to obtain the corresponding message.

The computational complexity of BP will be discussed in Section 4.3, as BP can be considered a especial case of the JT algorithm.

\subsection{Junction Tree}
In this section we describe the JT (or clique tree) algorithm, a method that takes the advantage of the message passing scheme of BP to compute all the marginals of a BN or a MRF, but is applied to general graphs rather than just trees and polytrees. JT aims to create a tree structure where the nodes represent clusters of the random variables in the graph, and then, apply message passing as in BP, to compute the unconditional probabilities. This method is equally applicable to both BNs and MRFs.

Besides presenting the BP algorithm for polytrees, \cite{pearl} also describes a simple approach to cluster nodes in general graphs. However, the technique produces very inefficient trees \cite{koller}. In contrast, the two variants of JT algorithm, namely the Shenoy-Shafer algorithm \cite{shenoy,shafer} and the Hugin algorithm \cite{lauritzen,jensen} produce more efficient trees by clustering nodes. Both techniques rely on the same principles although they differ in the way messages are computed. We will use here the Shenoy-Shafer method which uses the same message passing scheme as BP. A detailed comparison of Hugin and Shenoy-Shafer can be found in \cite{lepar}.


The first step of the JT algorithm is to create a cluster graph with a tree structure from the initial BN (or MRF). This can be viewed as an extension of factor graphs that clusters several random variables between two factors where each random variable can appear in more than one cluster node. The cluster or clique tree must also satisfy the running intersection property: if a random variable $X_i$ appears in two cluster nodes, $X_i \in C_j$ and $X_i \in C_k$, then, $X_i$ must also appear in each cluster node in the unique path existing between $C_j$ and $C_k$ in the clique tree.

As shown in \cite{koller}, an execution of VE induces a cluster graph with a tree structure that satisfies the running intersection property. Other procedures similarly rely on creating a chordal graph by moralizing and triangulating the original BN or MRF, so that each clique in the chordal graph is a cluster node in the clique tree. However, for the sake of clarity and since both solutions have a similar computational burden, we prefer to describe the use of VE as a method to obtain the clique tree.

\begin{figure}
	\centering
	\includegraphics[width=5.3cm,height=2.0cm]{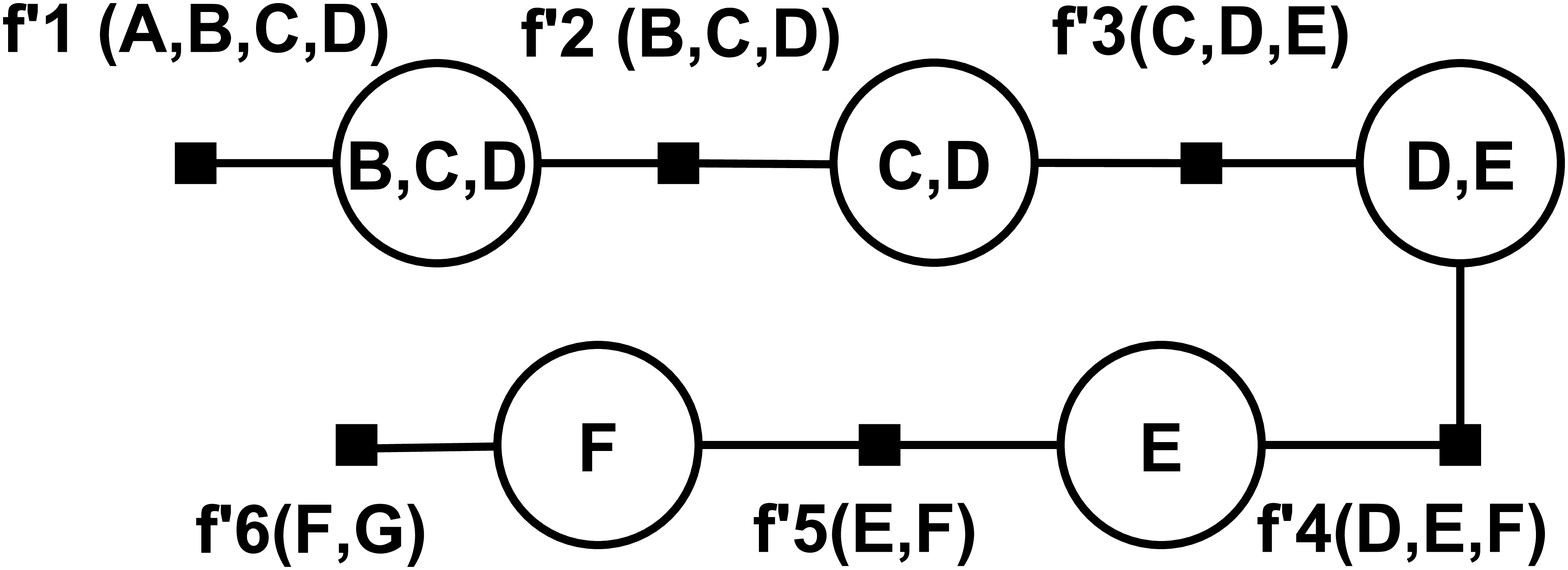} \\
	(a) \vspace{0.3cm} \\
	\includegraphics[width=7.3cm,height=1.3cm]{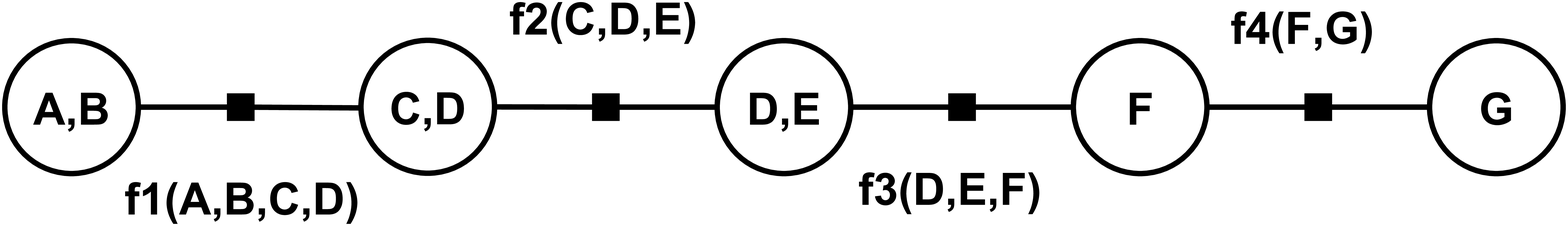} \\
	(b)
	\caption{(a) Initial factors for the JT of the BAG in Figure \ref{fig2} following the elimination order in Table \ref{tab2}. (b) Final factors for the JT after clustering factors and adding the leaf variable nodes.}
	\label{fig6}	
\end{figure}

To describe the procedure of generating the clique tree we use the BAG shown in Figure \ref{fig2} and the corresponding execution of VE following the elimination order ${\boldsymbol \Omega} = \{ A, B, C, D, E, F \}$ shown in Table \ref{tab2}. First, we create a initial factor $f'_i$ for each $\phi_i$ used in the computation of VE. Then, we draw an edge between $f'_i$ and $f'_j$ if the factor generated from $\tau_i$ is used in the computation of $\tau_j$. In our example, we have an edge between factors $f'_1(A,B,C,D)$ and $f'_2(B,C,D)$, corresponding to the terms $\phi_1(A,B,C,D)$ and $\phi_2(B,C,D)$ respectively  (because $\phi_2(B,C,D)$ depends on $\tau_1(B,C,D)$). Next, we add cluster nodes between each pair of factors considering all the random variables that intersect the two adjacent factors. For example, between $f'_1(A,B,C,D)$ and $f'_2(B,C,D)$ we include a cluster node with the variables $B,C,D$.  After these two steps, we get the JT shown in Figure \ref{fig6}.(a). However, we can reduce the tree by removing redundant factors (those whose scope is a subset of the scope of adjacent factors) removing also the corresponding cluster nodes between the implied factors. This is the case of factors $f'_2$ and $f'_5$ in our example, whose scopes are a subset of the scopes of $f'_1$ and $f'_4$ respectively. Finally, we add to each leaf factor node a cluster node, which becomes a leaf node, with the random variables that are in the scope of that factor but are not in the other cluster nodes associated to it. For example, for $f'_1$ we should add a cluster node with the variables $A$ and $B$, since the other cluster node associated to $f'_1$ contains $C$ and $D$ and the scope of $f'_1$ is $\{ A,B,C,D \}$. The reduced final clique tree is shown in Figure \ref{fig6}.(b).

Once we have defined the factors and obtained the reduced JT, as in the case of BP, we associate each $f_i$ with different factors from the joint probability distribution. This can be done by simply assigning the set of factors used at each step in VE to the corresponding factor in the clique tree. Thus, in our example, one possible assignment is:
\begin{equation}
\begin{split}
f_1(A, B, C, D) & = p(A) \ p(B|A) \ p(C|A, B) \ p(D|A) \\
f_2(C, D, E) & = p(E|C) \\
f_3(D, E, F) & = p(F|D, E) \\
f_4(F, G) & = p(G|F) \\
\end{split}
\label{jt2}
\end{equation}

To calculate the unconditional probabilities in the JT we use the same message passing scheme as in BP. The only difference is that the scopes of the messages from nodes to factors and from factors to nodes, given in equations (\ref{bp3}) and (\ref{bp4}), depend on multiple random variables rather than just one. For example, the scope of the message from cluster node $A,B$ to $f_1$ is $A,B$, i.e. $\mu_{AB,f_1} (A,B)$. The list of all messages needed to calculate the marginal probabilities are shown in the Supplementary Material.

The unconditional joint probability for the variables in cluster node ${\bf X}_s$, assuming the graph is a BN, is given by:
\begin{equation}
p({\bf X}_s) = \prod_{f_j \in {\bf F}_s} \mu_{f_j,{\bf X}_s}({\bf X}_s)
\label{jt3}
\end{equation} where ${\bf F}_s$ are the factor nodes in the neighbourhood of the cluster node ${\bf X}_s$. To calculate the marginal probability for one random variable in the set ${\bf X}_s$ from $p({\bf X}_s)$, we just sum over the other variables in ${\bf X}_s$. For example, the expression to calculate $p(C)$, the probability that an attacker obtains user privileges in \emph{Host 1}, is given by:
\begin{equation}
p(C) = \sum_D p(C,D) = \sum_D \mu_{f_1,CD} \cdot \mu_{f_2,CD}
\label{jt4}
\end{equation} 

Evidence of compromise can then be included in inferences using the JT algorithm in the same way as in BP, as previously described. 

Finally, as in the case of VE, the computational complexity of the JT algorithm is exponential in the scope of the biggest factor in the clique tree. Concretely, if all variables in the graph are discrete and have $K$ possible values each (in our case, $K = 2$), JT scales in time and space as ${\cal O} (|F| K^s)$, where $|F|$ is the number of factors and $s$ is the size of the scope of the largest factor in the clique tree. JT therefore suffers from the same scalability problems as VE. However, for JT, we only need to run VE to create the clique tree once and then compute and store all the messages, which is significantly more efficient than VE, where the algorithm needs to be run from the start each time we want to compute the marginal probability for a single node. 

\section{Experiments}
In this section we present our experimental results comparing the  performance of JT to that of the VE algorithm used in \cite{liu}. More broadly, the purpose of the experiments is to analyse the performance of the VE and JT algorithms for inference in BAGs in terms of both time and the memory requirements. This allows us to examine their suitability for static and dynamic analysis of AGs, as the BAG models proposed in the literature \cite{frigaultMsC,liu,poolsappasit} have not been evaluated experimentally. Beyond the differences between VE and JT when used for inference in BNs, we also want to highlight, through these experiments, the impact of the elimination ordering heuristic. For VE we establish the elimination ordering at random, since in \cite{liu} no elimination ordering is proposed, whereas for JT, we use the min-weight heuristic \cite{koller} to find the elimination ordering needed to build the clique tree. For the implementation of both algorithms we use the Bayes Net toolbox for Matlab\footnote{Implementation available at \url{https://github.com/bayesnet/bnt}\\ Our code implementation with the experiments for the BAG model can be found at \url{http://rissgroup.org/}}. 

To provide a comprehensive evaluation of the algorithms we have used synthetic AGs, as many graphs of different sizes and structure are needed to provide meaningful performance results. Such a broad range of empirical AGs is not available especially as generating AGs for real systems is far from trivial. Furthermore, typical structures for AGs are not clear from the examples encountered in the literature, and we expect them to vary significantly since they depend on both the network topology and the number of vulnerabilities. In view of these limitations and in order to give a good characterisation of the performance of VE and JT we have considered two kind of structures for our synthetic AGs: \textit{Random graphs}, where we control the in-degree of the nodes (which is related to the number of vulnerabilities on each network node), and \textit{cluster graphs}, where we explore the behaviour of the algorithms with respect to the size of the clusters. We expect real use-cases to have a range of structures between these two kind of graphs depending on the network topology. We also include in our discussion an AG generated using a realistic use-case representative of the corporate network of an SME.

The values for the probabilities of exploitation of vulnerabilities needed to build the conditional probability tables are drawn at random, as well as the kind of logical conditions to build these tables: AND/OR. It is important to note that this does not have an impact on the time or memory required to calculate the unconditional probabilities with VE and JT, but only on the values of the probabilities of the nodes. 

\subsection{AG example}
For our first example, we have built a realistic small use-case scenario, to show the results of our proposed techniques in a understandable way. Later, we will show more tests performed on larger AGs generated synthetically.

\begin{figure}
	\centering
	\includegraphics[width=9.0cm]{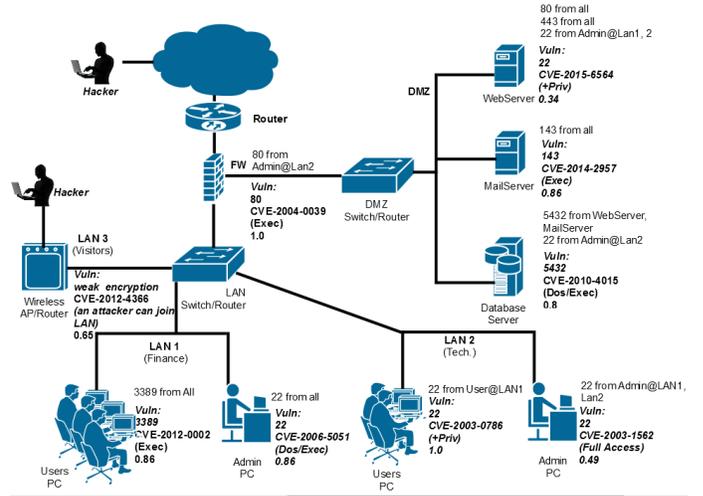} \\
	\caption{Example of a network representative of an SME.}
	\label{exampleNet}	
\end{figure}

Figure \ref{exampleNet} shows a typical network for an SME. In detail, we have two internal LANs (one for finance/accounting, one for technicians), a Wireless LAN for visitors (but that, if compromised, can be also used to reach the internal network), and finally a DMZ hosting the SME servers (in the example, a public Web Server, a Mail Server, and a Local Database used to store public and private data). In Figure \ref{exampleNet} we show for each node its reachable ports and the nodes from where they are reachable (this includes those ports open/filtered by the firewall as well as those open/closed by local firewalls, switches, routers, etc.). Additionally, we have highlighted vulnerabilities that may be present on the nodes, the port on which they can be exploited (if remote), their CVE identifier, the type of vulnerability (DoS, elevation of privilege, etc.), and the likelihood of exploiting it. We have based this likelihood on the CVSS Exploitability Subscore (divided by 10), which is in line with existing use in the literature. Although, as discussed in Section 3, this fails to account for the resources/skills of the attackers and the knowledge of the existence of an exploit among others. In our example, when the CVSS Exploitability Subscore is 1.0 (i.e., an easy to use exploit already exists), we have lowered its value to 0.95. A probability of successful exploitation of 1.0 would mean that the attacker has already reached the next security state without necessarily exploiting the vulnerability, which is not true.

\begin{figure}
	\centering
	\includegraphics[width=7.5cm,height=6.2cm]{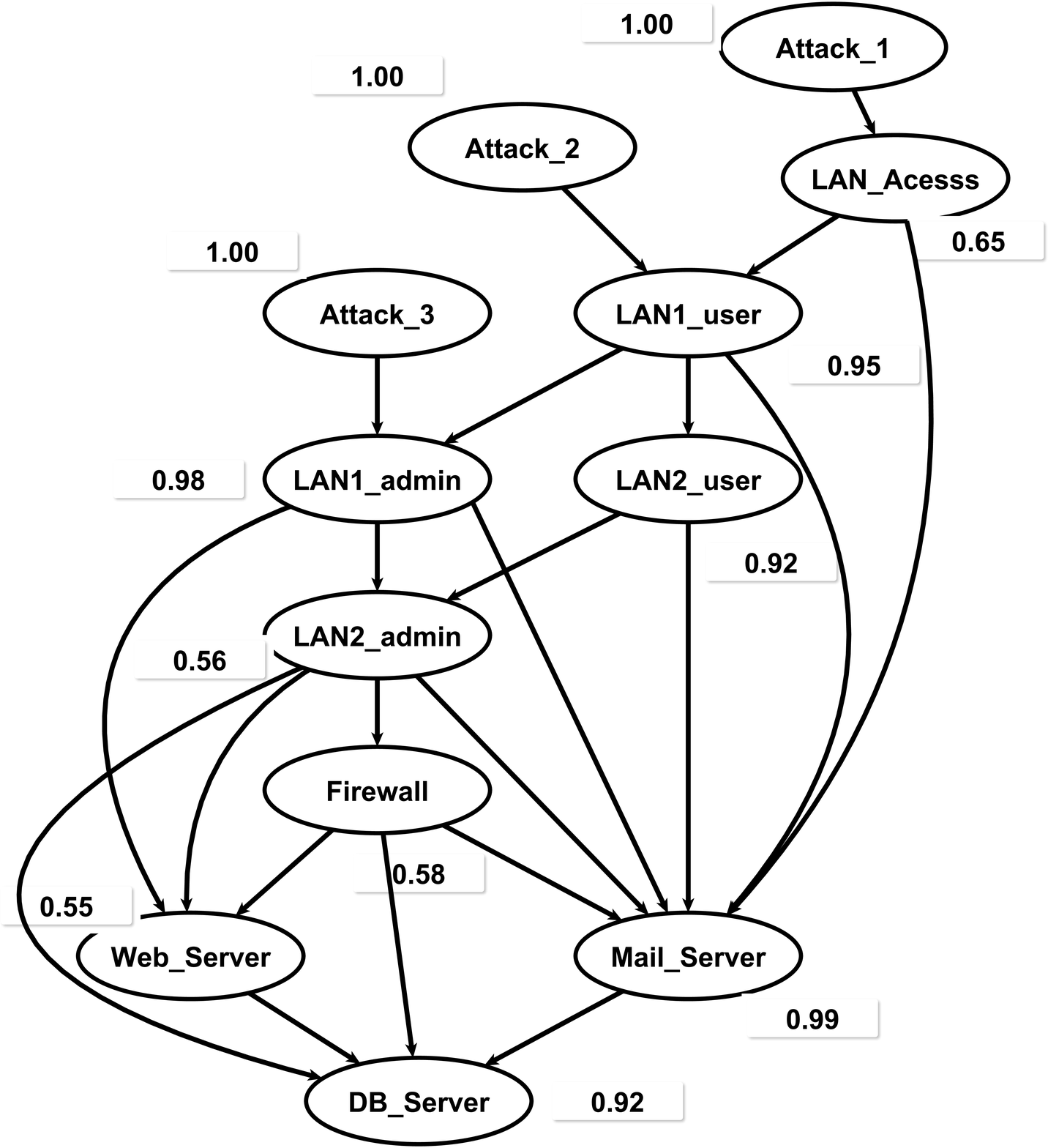} \\
	\caption{BAG for the network in Figure \ref{exampleNet} showing the unconditional probabilities that an attacker can successfully reach a security state for all the nodes in the BAG.}
	\label{BAGexampleNet}	
\end{figure}

In Figure \ref{BAGexampleNet} we show the BAG obtained when the goal of the attacker is assumed to be to compromise the Database Server. Note that we have clustered all (similar) users in one single node in the BAG, and similarly for the system administrators. From an AG perspective, compromising one or several machines with the same behaviour (in terms of connectivity, services running, and vulnerabilities) can be considered the same in terms of privileges acquired towards compromising a given target. In contrast, the number of similar machines compromised is important when considering more generic targets such as information leakage or botnet recruitment. Furthermore, although in our example clustering similar machines is a simplification of the reality (we assume their behaviour to be exactly the same, which is not necessarily the case) the approximation still depicts in quite a realistic way the kind of AG we can produce for typical corporate networks. In our example, we do not consider insider attackers, and we assume that attackers start the attack from the Internet or from the visitors Wireless LAN. 

Using the JT algorithm, the time required to create the clique tree and to compute the marginal probabilities for each node in the BAG, shown in Figure \ref{BAGexampleNet}, is less than 0.05 seconds on a common desktop computer, which makes JT suitable as a tool for the analysis of this AG. Although this example intends to be representative of a small corporate network, in some practical situations we can expect more complex and much larger AGs. To assess the suitability of JT for the analysis of AGs, we use, in the rest of the experiments, synthetic AGs with varying numbers of nodes and topologies.

\subsection{AGs with pseudo-random structure}
For this kind of graphs we build the BAG directly by generating random DAGs where we limit the maximum number of parents for each node. This corresponds to limiting the maximum number of vulnerabilities that can allow an attacker to obtain a certain security condition. In our opinion, this assumption is reasonable as on real (managed) corporate networks we expect to have a limited number of vulnerabilities on each host. Then, for each node $X_j$ in the graph, we randomly select the number of parents by drawing an random integer $n_p$ in the interval $[1, m]$ uniformly, where $m$ is the maximum number of parents. After that, we randomly select $n_p$ parent nodes for $X_j$ from the set of nodes in the BAG for which $X_j$ is not a parent node (to avoid directed cycles). For $m$, we have explored the values $2$, $3$, and $4$. These values are selected on the assumption that the number of vulnerabilities to reach a security state is expected to be reduced. For the number of nodes $n$ in the graph we have explored values in the range $[10, 220]$. However, depending on the algorithm and the value of $m$ we have limited the value of $n$ because of physical memory limitations. For example, VE could only be applied to random networks with $m=4$ up to $n = 60$ nodes. For each value of $n$ and $m$, we have generated $20$ random networks and, for each network, we compute the unconditional probabilities for all the nodes with VE and JT. 

\begin{figure}
	\centering
	\psfrag{VE-2}{{\scriptsize VE-2}}
	\psfrag{VE-3}{{\scriptsize VE-3}}
	\psfrag{VE-4}{{\scriptsize VE-4}}
	\psfrag{JT-2}{{\scriptsize JT-2}}
	\psfrag{JT-3}{{\scriptsize JT-3}}
	\psfrag{JT-4}{{\scriptsize JT-4}}
	\psfrag{Number of nodes}[][]{{\scriptsize Number of nodes}}
	\psfrag{Time (s)}[0.5cm][-0.5cm]{{\scriptsize Time (s)}} 
	\includegraphics[width=7.8cm,height=5.0cm]{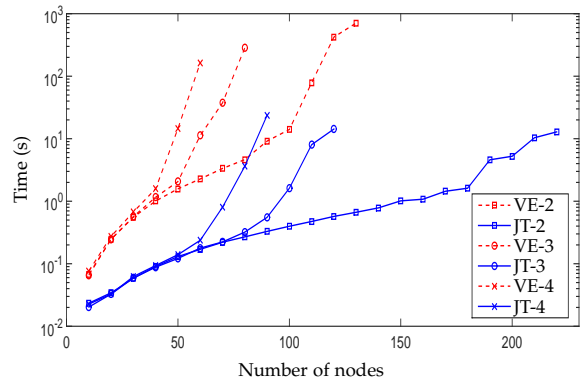}
	\caption{Average time to compute the unconditional probabilities for VE and JT. The notation VE-$m$ and JT-$m$ stands for the value of $m$, the maximum number of parents of each node, used to generate the graphs in each case.}
	\label{res1}	
\end{figure}

We show in Figure \ref{res1} the average time required to compute all the unconditional probabilities for VE and JT. In the case of JT, the reported time includes the time required to find the elimination order, build the clique tree, compute all the messages, and calculate the marginal probabilities. The difference in performance between the algorithms is remarkable: JT is much faster than VE in all the cases considered. For example, for $n = 130$ and $m = 2$, the average time to compute the unconditional probabilities is less than 1 second for JT, whereas it is almost 1000 seconds for VE. As described in Section 4, JT only needs to build the clique tree and compute the messages once, while in VE we compute everything each time we want to calculate the unconditional probability in one network node, thus much less efficient. Besides, the elimination ordering also plays an important role: a good elimination ordering can reduce the scope of the factors appearing in the clique tree (or the $\phi_i$ factors in VE), and thus reduces the size of the tables that need to be computed and requires less memory. 

\begin{figure}
	\centering
	\psfrag{VE-2}{{\scriptsize VE-2}}
	\psfrag{VE-3}{{\scriptsize VE-3}}
	\psfrag{VE-4}{{\scriptsize VE-4}}
	\psfrag{JT-2}{{\scriptsize JT-2}}
	\psfrag{JT-3}{{\scriptsize JT-3}}
	\psfrag{JT-4}{{\scriptsize JT-4}}
	\psfrag{Base}{{\scriptsize Baseline}}
	\psfrag{Number of nodes}[][]{{\scriptsize Number of nodes}}
	\psfrag{Avg. max number of clustered nodes}[0.5cm][-0.5cm]{{\scriptsize Avg. max \# of clustered nodes}} 
	\includegraphics[width=7.8cm,height=5.0cm]{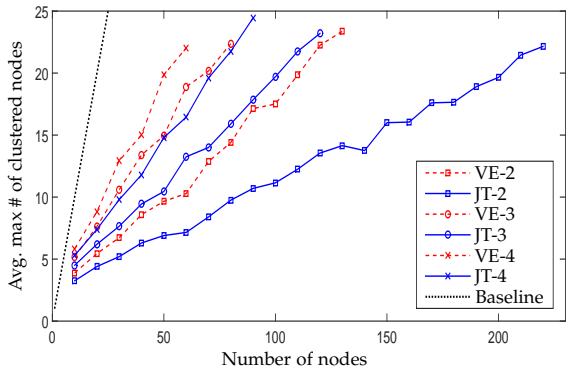}
	\caption{Average number of nodes of the biggest factor for VE and JT. The notation VE-$m$ and JT-$m$ stands for the value of $m$, the maximum number of parents of each node, used to generate the graphs in each case. The black-dotted line indicates the memory required when applying brute force.}
	\label{res2}	
\end{figure}

We report in Figure \ref{res2} the average number of nodes in the scope of the biggest factor for both VE and JT, as a function of the number of nodes. This corresponds to the number of variables of the biggest $\phi_i$ and $f_i$ for VE and JT respectively. We first note the big computational savings of VE and JT when compared to applying brute force, i.e. computing the joint probability on all the nodes. We further appreciate, that, again, JT is more efficient than VE in all the cases. However, the difference between them here is only due to the elimination ordering algorithm. As mentioned before, we have not used any elimination ordering for VE, as none was proposed in \cite{liu}, whereas for JT we have used the min-weight heuristic. When using the same elimination ordering, we should expect similar results for VE and JT. We can observe the computational savings in terms of memory (and then, also in time) when using an elimination ordering heuristic. Note that the tables needed to compute the factors in VE and JT grow exponentially with the number of variables in the scope of the factors. For example, for $n = 130$ and $m = 2$ the average number of nodes in the biggest factor for VE is approximately $23$, whereas for JT is $14$. This means that the average memory required to store these factors (considering that we use 8 bytes to store each entry in the table) is $2^{23 + 3} / 1024^2 = 64$ Megabytes for VE without an elimination ordering heuristic and only $2^{14 + 3} / 1024^2 = 0.125$ Megabytes for JT using the min-weight heuristic.

\begin{figure}
	\centering
	\psfrag{Number of nodes}[][]{{\scriptsize Number of nodes}}
	\psfrag{Time (s)}[0.5cm][-0.5cm]{{\scriptsize Time (s)}} 
	\psfrag{Static} {{\scriptsize Static}}
	\psfrag{Dynamic} {{\scriptsize Dynamic}}
	\includegraphics[width=7.9cm,height=4.2cm]{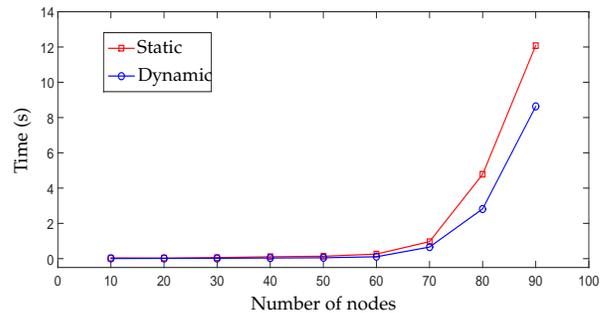}
	\caption{Time to compute the unconditional probabilities with the JT for pseudo-random graphs with $m = 4$ for the static and the dynamic analysis of the BAGs (when we observe evidence at one node).}
	\label{res3}	
\end{figure}

As described in Section 4, when evidence of compromise is observed in some of the graph nodes, the JT algorithm does not need to build the clique tree again but only to recompute the messages taking into account the evidence. In this sense, \textit{static analysis} with the JT algorithm consists in building the clique tree, computing all the messages and calculating the unconditional probabilities in the absence of any evidence of compromise. In contrast, \textit{dynamic analysis} consists only in recomputing all the messages and recalculating the marginal probabilities (conditioned on the evidence). This scenario is frequent in AG analysis when correlating with compromise evidence from IDS and SIEM at run-time. Figure \ref{res3} shows the difference in time between the static and the dynamic analysis using the JT for graphs with $m = 4$ and nodes $n$ varying from 10 to 90. For the dynamic analysis we randomly select one node where we consider the evidence has been observed. It is interesting to note that, although the time to recompute the probabilities is lower, the difference is not very significant. This means that, for pseudo-random graphs, the bottleneck of the JT algorithm is the computation of the messages rather than the elimination ordering algorithm used to build the tree, since the number of variables for each factor is high (at least for some factors), as we can observe in Figure \ref{res2}. This is due to the fact that the generated graphs are highly connected. This observation does not apply in the case of VE as the steps to compute the marginal probabilities are the same for static and dynamic analysis.

\subsection{AGs with cluster structure}
\begin{figure}
	\centering
	\psfrag{Number of nodes}[][]{{\scriptsize Number of nodes}}
	\psfrag{Time (s)}[0.5cm][-0.5cm]{{\scriptsize Time (s)}} 
	\psfrag{VE} {{\scriptsize VE}}
	\psfrag{JT} {{\scriptsize JT}}
	\includegraphics[width=7.3cm,height=4.0cm]{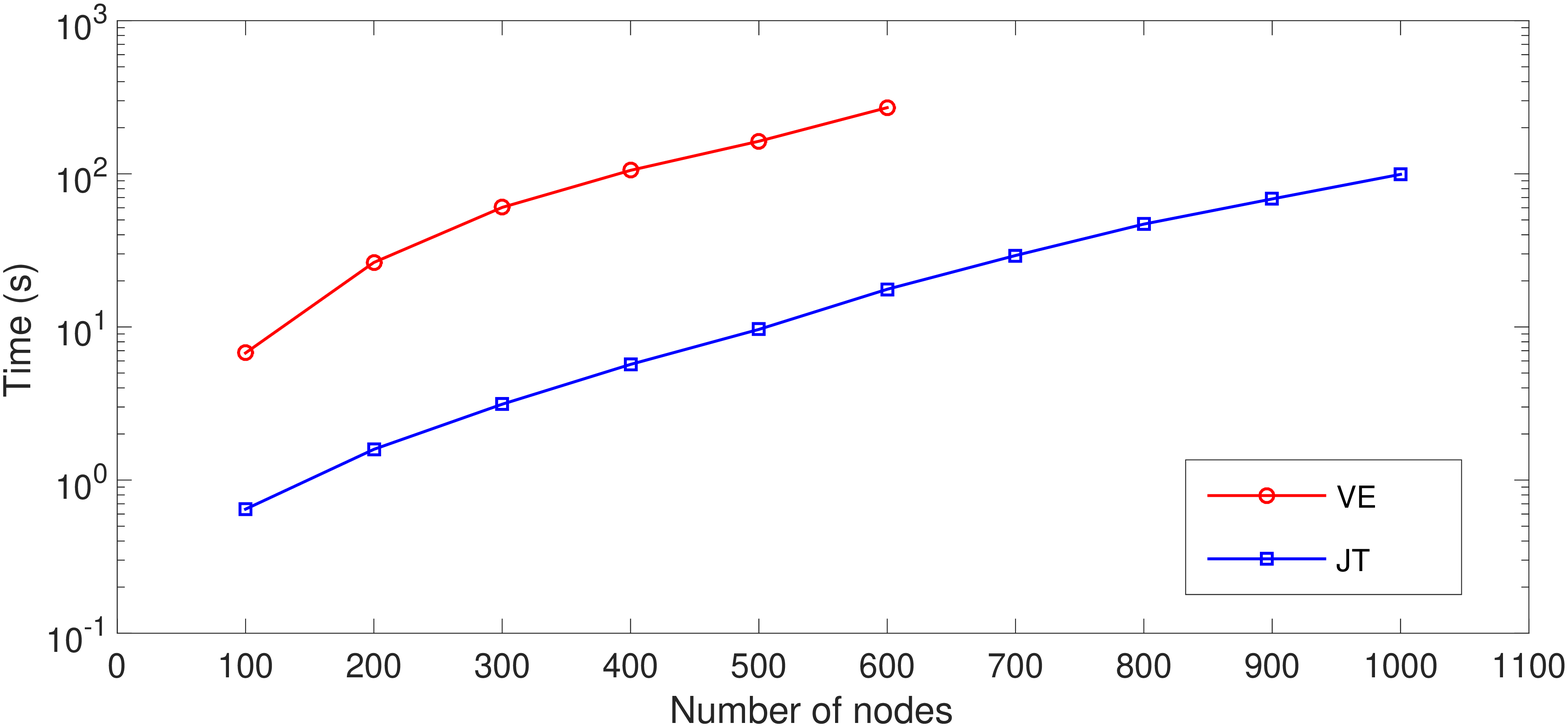} \\
	(a) \vspace{0.3cm} \\
	\includegraphics[width=7.3cm,height=4.0cm]{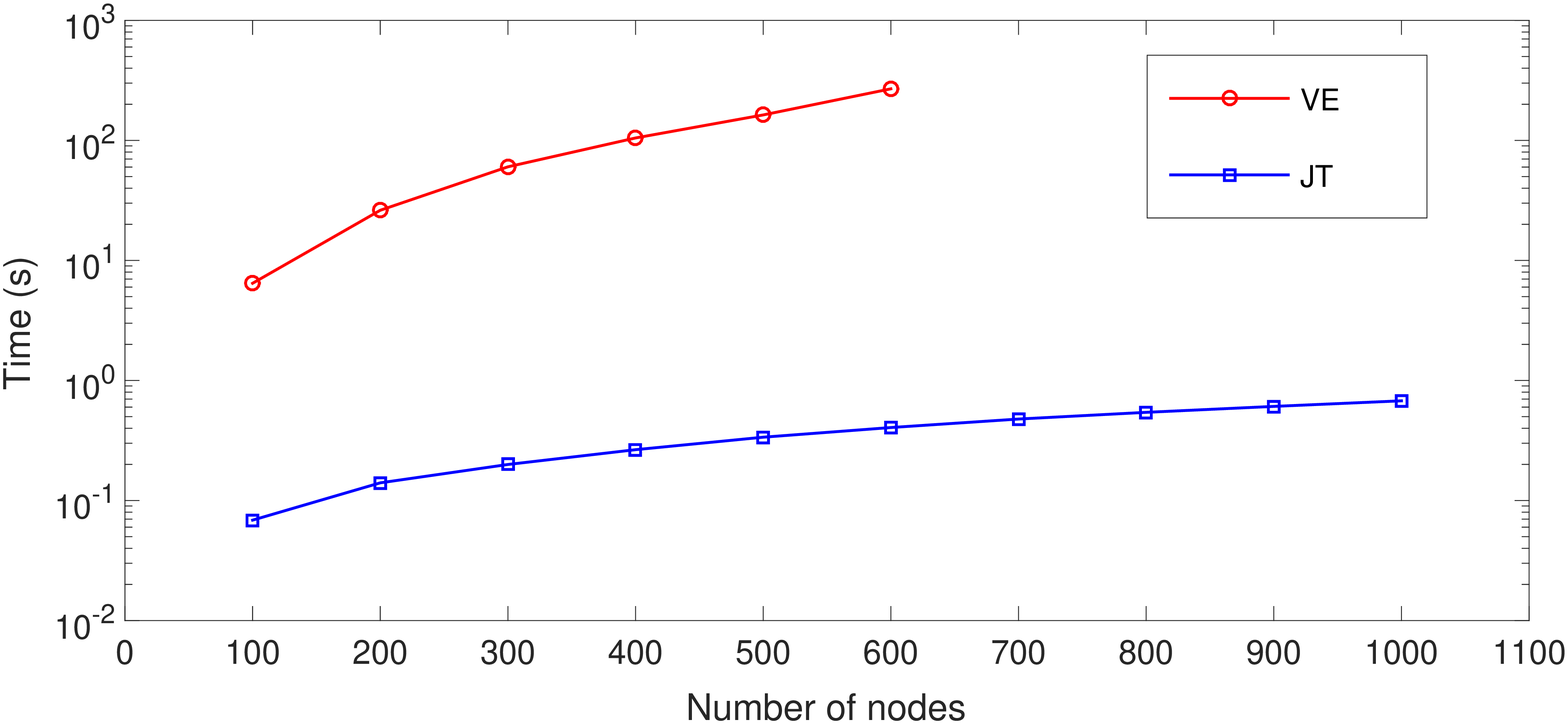} \\
	(b)  \\
	\caption{Time to compute the unconditional probabilities for VE and JT for cluster BAGs with clusters of size $10$ and $m = 4$: (a) for the static analysis and (b) for the dynamic analysis (when we observe evidence at one node).}
	\label{res4}	
\end{figure}

To generate synthetic graphs with a cluster structure, we have considered networks with clusters of the same size, $n_c$. For each cluster, we have generated pseudo-random subgraphs limiting the maximum number of parents for each node to $m$. Finally, we have included dependencies between clusters by adding one edge from one node in each cluster to one node in other clusters provided that the added edge preserves the DAG structure required for BNs. For the first experiment we have set $n_c = 10$ and varied the nodes in the network from $100$ to $1000$. Then, we have measured the time required to compute the unconditional probabilities using VE (with random elimination ordering) and JT (using min-weight to build the clique tree) for both the static and the dynamic analysis, considering that we observe evidence at one node. The results in Figure \ref{res4} show again a significant difference in performance between JT and VE, as in the case of pseudo-random graphs. The difference here is mainly due to the fact that in VE we need to recompute everything each time we calculate the marginal probability for a node rather than the lack of an elimination ordering heuristic for VE.

We can observe in Figure \ref{res4} that the time required by VE to calculate the marginal probabilities is almost the same for the static and the dynamic analysis. This is not surprising as VE basically performs the same operations in both. In contrast, JT shows a noticeable difference between the time required for the static and the dynamic analysis. In this case, the proportion of time required to build the clique tree is significantly higher than the proportion of time needed to compute all the messages and calculate the probabilities. Note that this behaviour is just the opposite of that observed for pseudo-random networks and can be explained by the clustered structure of the graph, which leads to smaller factors in the clique tree. 

\begin{figure}
	\centering
	\psfrag{VE}{{\scriptsize VE}}
	\psfrag{JT}{{\scriptsize JT}}
	\psfrag{Base}{{\scriptsize Baseline}}
	\psfrag{Number of nodes}[][]{{\scriptsize Number of nodes}}
	\psfrag{Number of clustered nodes}[0.5cm][-0.5cm]{{\scriptsize Avg. max \# of clustered nodes}} 
	\includegraphics[width=7.3cm,height=4.0cm]{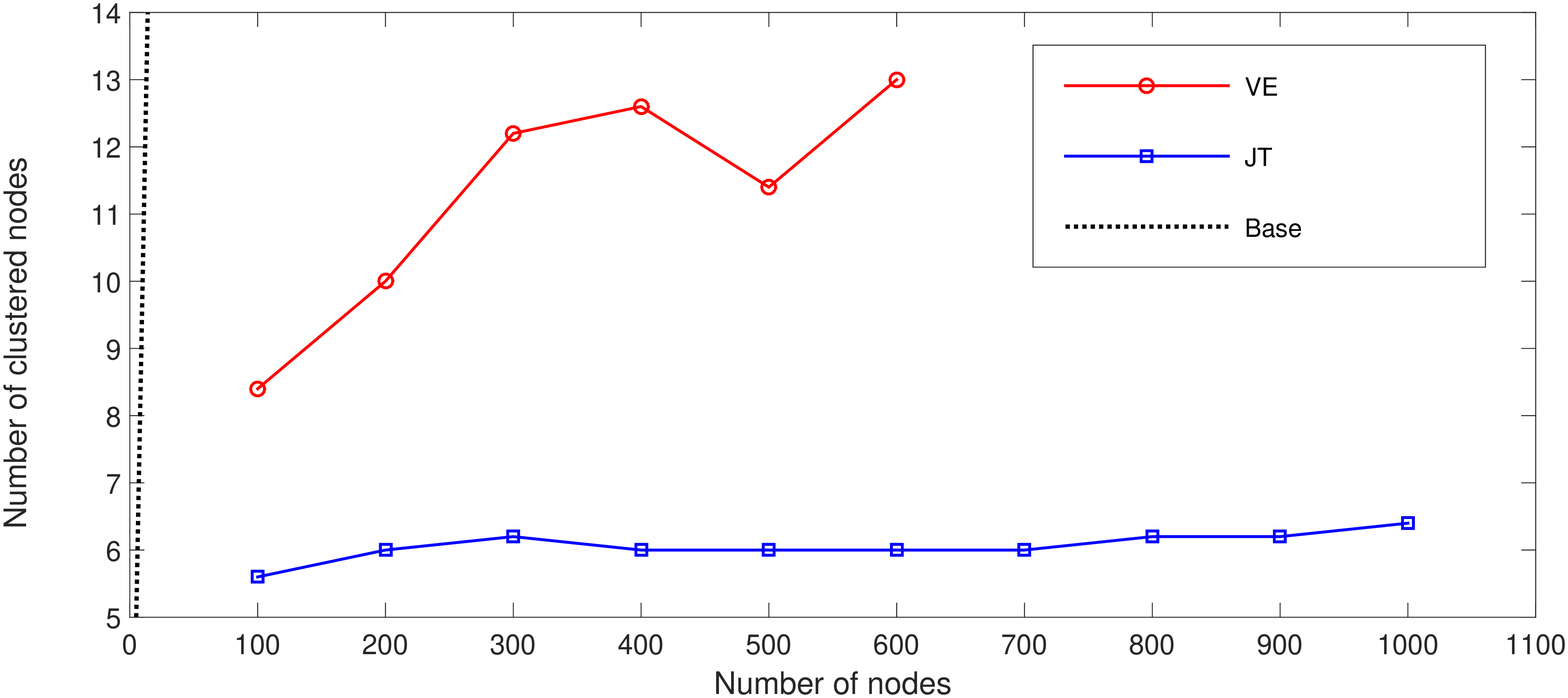}
	\caption{Average number of nodes of the biggest factor for VE and JT for the cluster BAGs. The black-dotted line indicates the memory required when applying brute force.}
	\label{resMem}
\end{figure}

In Figure \ref{resMem} we show the average number of nodes in the scope of the biggest factor for VE and JT. Note that, again, the difference of applying VE or JT with respect to the baseline (applying brute force) is huge. Also, as in the case of the pseudo-random BAGs, we observe a signifcant difference between JT and VE. In particular, the average number of nodes in the scope of the biggest factor increases slowly with the number of nodes in the graph for JT, in contrast to VE. As before, these differences can be explained by the use of the min-weight heuristic to find an elimination ordering when using JT. 

\begin{figure}
	\centering
	\psfrag{Number of nodes}[][]{{\scriptsize Number of nodes}}
	\psfrag{Time (s)}[0.5cm][-0.5cm]{{\scriptsize Time (s)}} 
	\psfrag{JT-10} {{\scriptsize JT-10}}
	\psfrag{JT-30} {{\scriptsize JT-30}}
	\psfrag{JT-50} {{\scriptsize JT-50}}
	\includegraphics[width=7.3cm,height=4.0cm]{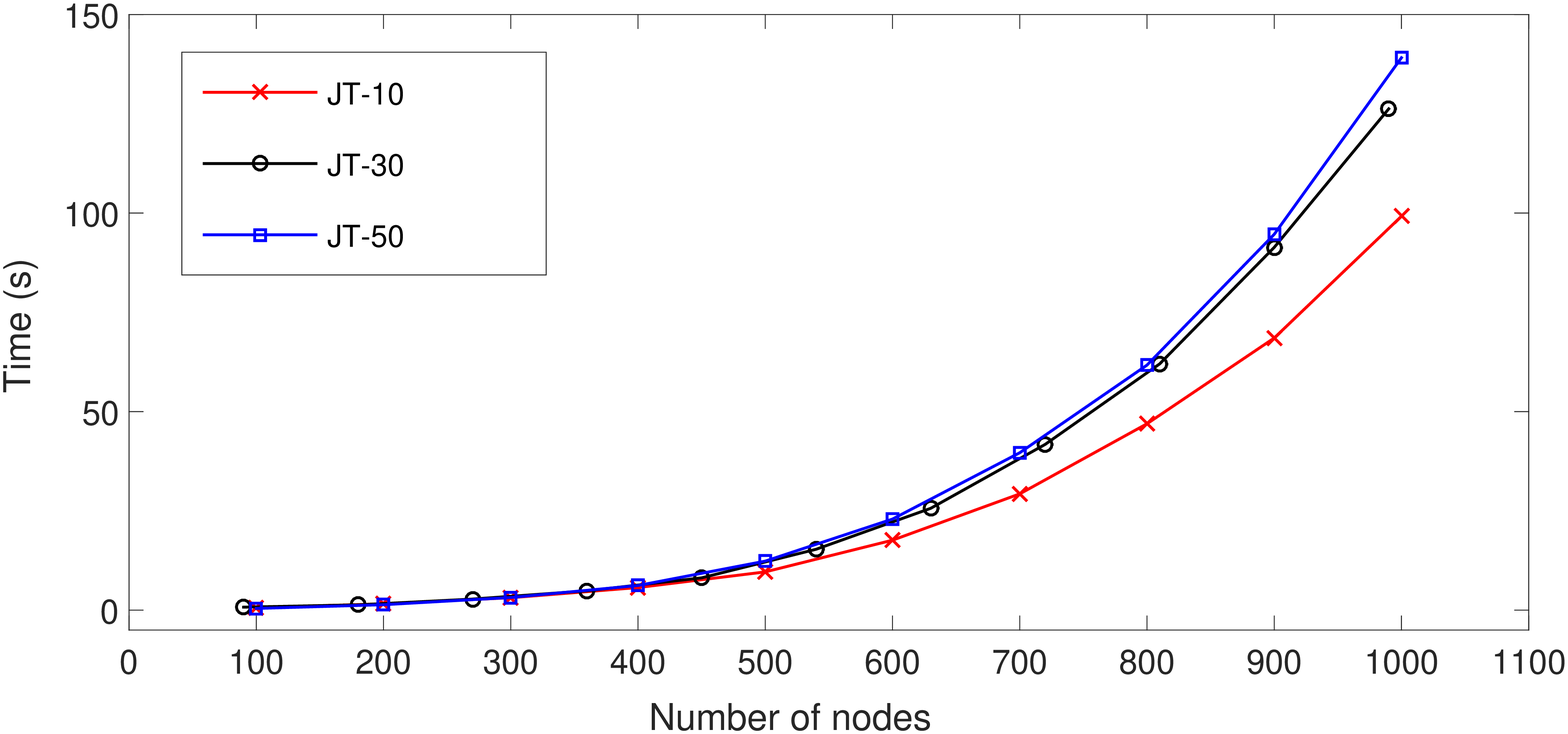} \\
	(a) \vspace{0.3cm} \\
	\includegraphics[width=7.3cm,height=4.0cm]{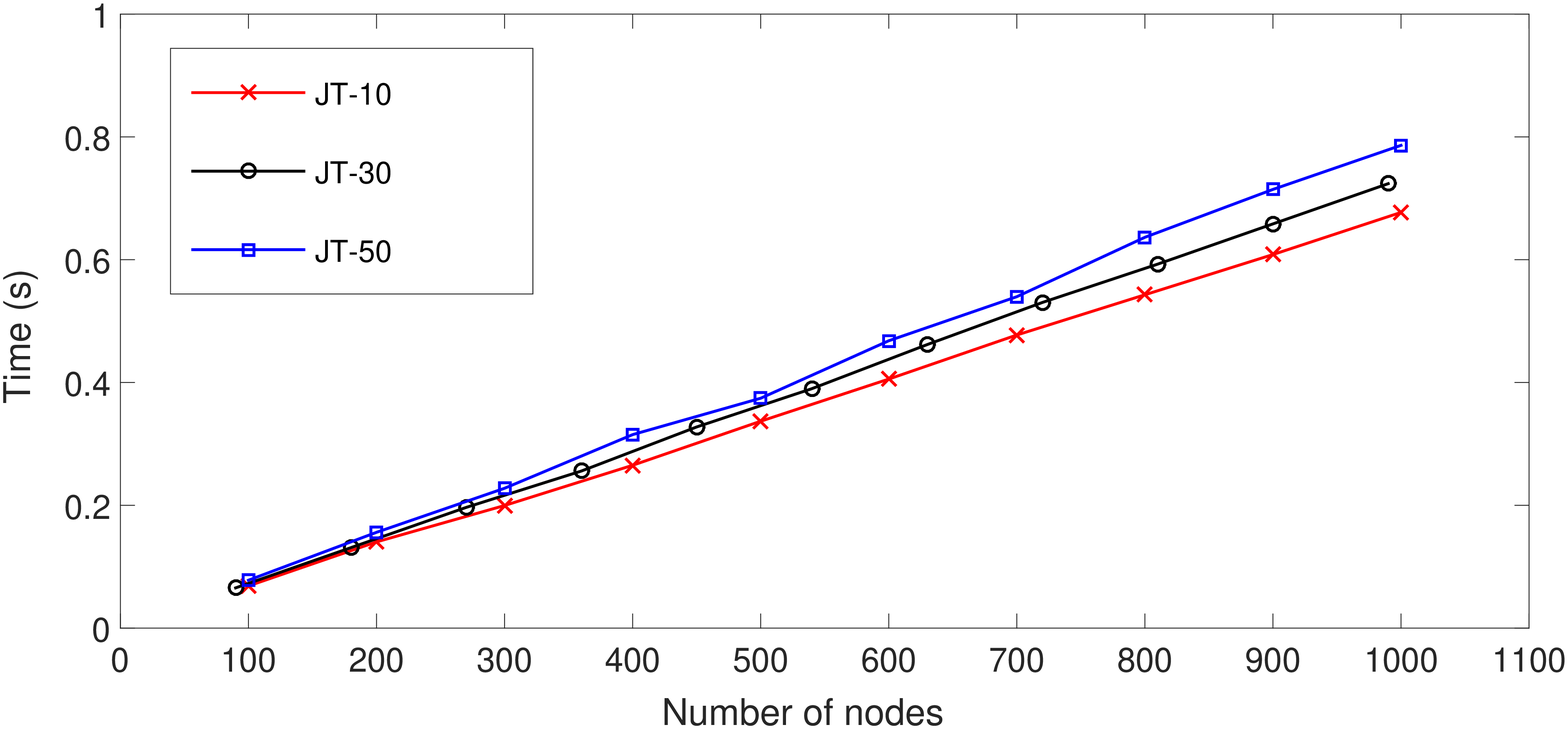} \\
	(b)  \\
	\caption{Time to compute the unconditional probabilities for JT algorithm for cluster BAGs with different cluster sizes ($10, 30$, and $50$) and $m = 4$: (a) for the static analysis and (b) for the dynamic analysis (when we observe evidence at one node).}
	\label{res5}	
\end{figure}

In Figure \ref{res5} we show the performance of JT for the static and the dynamic analysis in cluster graphs whilst varying the size of the clusters (from $10$ to $50$) and the size of the graph (from $100$ to $1000$), setting $m=4$. As in Figure \ref{res4}, the difference in time between the two analyses is significant. Moreover we can appreciate a different behaviour: Whilst the time required for static analysis still scales exponentially with the number of nodes, that required for dynamic analysis broadly scales linearly. The exponential behaviour of the static analysis is due to the elimination ordering algorithm used to build the clique tree. However, the clustered structure of the graph combined with weak dependencies between clusters result in factors of a similar size regardless of the network size. This allows the computation of the messages and of the probabilities to scale nearly linearly.

\begin{figure}
	\centering
	\psfrag{JT-10} {{\scriptsize JT-10}}
	\psfrag{JT-30} {{\scriptsize JT-30}}
	\psfrag{JT-50} {{\scriptsize JT-50}}
	\psfrag{Number of nodes}[][]{{\scriptsize Number of nodes}}
	\psfrag{Number of clustered nodes}[0.5cm][-0.5cm]{{\scriptsize Avg. max \# of clustered nodes}} 
	\includegraphics[width=7.3cm,height=4.0cm]{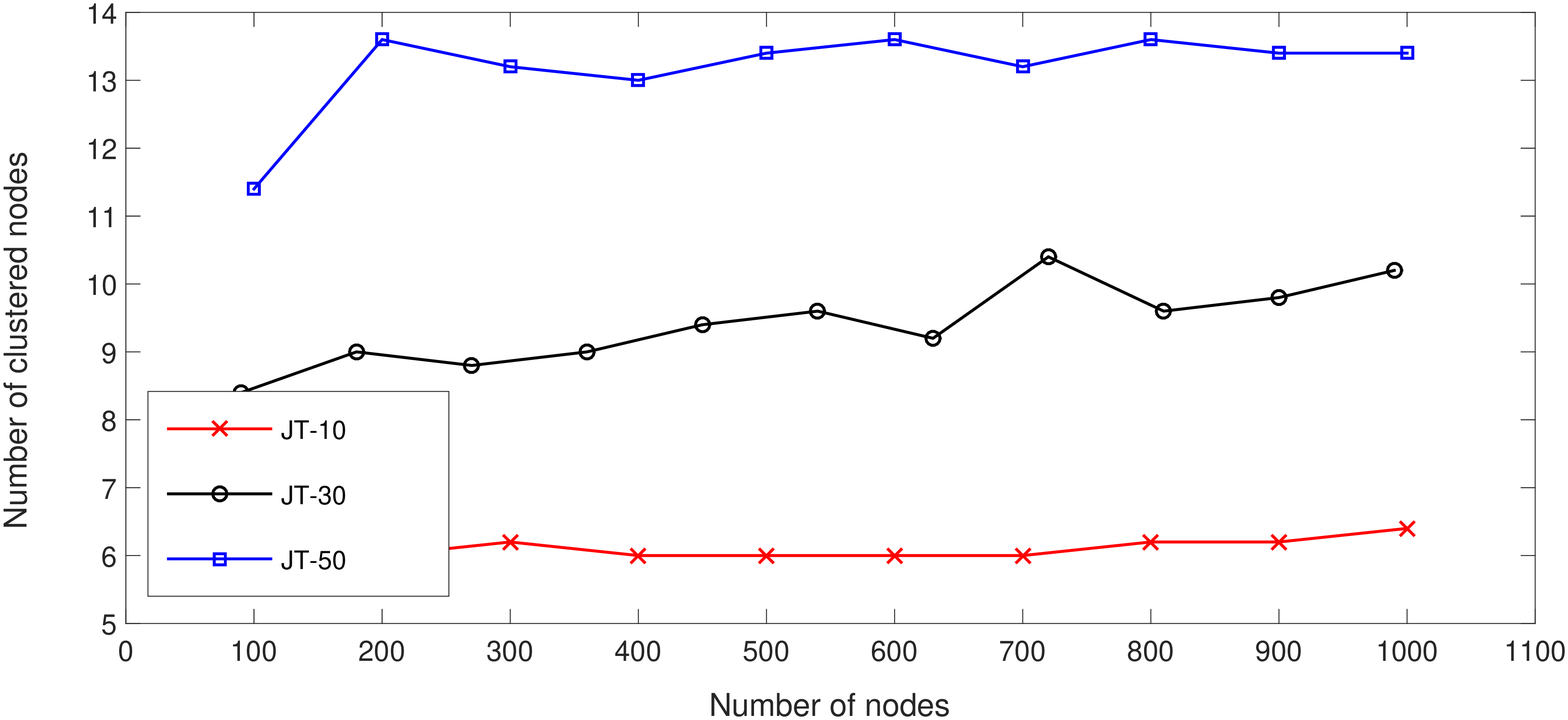}
	\caption{Average number of nodes of the biggest factor for VE and JT for the cluster BAGs. The black-dotted line indicates the memory required when applying brute force.}
	\label{resMemJT}	
\end{figure}

It is important to highlight the reduced time (less than 1 s) required by JT to recompute the unconditional probabilities when we observe evidence of compromise. This makes JT an appealing choice to perform dynamic analysis on real AGs, to select and apply risk mitigation strategies at run-time. In contrast, the high cost of recomputing the same probabilities for VE renders it unusable in practice. For static analysis, the exponential memory and time scalability of JT (and so, VE) still limits their application to medium-size graphs, although this analysis can be done off-line and also depends on the topology of the AG. Furthermore, modelling with AGs can be performed at different levels of granularity. For example, the AG model in \cite{cauldron} only considers vulnerabilities that let an attacker move from one subnetwork to another. From the AG perspective this means that compromises within a subnetwork are considered equivalent for the risk assessment of the whole network. In this sense, intermediate levels of granularity can be adopted to achieve better accuracy when modelling the AG. In any case, taking into account the cluster structure of networks in AG modelling is a key aspect to be considered to perform a tractable dynamic risk assessment of networks with BAGs using the JT algorithm.

In Figure \ref{resMemJT} we show the average number of nodes in the scope of the biggest factor as a function of the total nodes when applying JT and varying cluster size. Similarly to Figure \ref{resMem}, we observe that the increase in size of the biggest factor is slow. Thus, the factor that affects most the memory required to apply JT is the size of the clusters rather than the total number of nodes in the graph.

\section{Extensions of the BAG Model}

The BAG model presented in Section 3 can be extended to analyse more complex aspects of an attack.

The effect of zero-day vulnerabilities can be considered by adding one extra node (with no parents) connected to each of the security condition nodes in the graph. The difficulty lies in estimating a reasonable score to rate their probability of successful exploitation when building the conditional probability tables. In \cite{poolsappasit} the network administrator is required to quantify it, but this is not a trivial task. The same reasoning applies when considering insider or social engineering attacks.

To model the uncertainty about the evidence that a node may be compromised (based on IDS and SIEM information) we should include correction terms in equations (\ref{eq4}) and (\ref{eq5}). One possible solution is to add one extra parent node for each existing node in the BAG, setting the prior probabilities of these added nodes to $1$ (as in the case of the attacker's initial state node). Then, we can use the error probability of the IDS as a parameter to build the conditional probability tables of the corresponding nodes. The modification of (\ref{eq4}) to compute the probability table in the AND case including this uncertainty results in:
\begin{equation}
p(X_i | {\bf pa}_i) = \begin{cases} p_e, & \exists X_j \in {\bf pa}_i | X_j = F \\ \begin{split} & 1 - (1 - p_e) \times \\ & \ (1 - \prod_{j: X_j} p_{v_j}) \end{split}, & \text{otherwise} \end{cases}
\label{ext1}
\end{equation} In the OR case, the modification of expression (\ref{eq5}) is given by:
\begin{equation}
p(X_i | {\bf pa}_i) = \begin{cases} p_e, & \forall X_j \in {\bf pa}_i | X_j = F \\ \begin{split} & 1 - (1 - p_e) \times \\ & \ \prod_{j: X_j} (1 - p_{v_j}) \end{split}, & \text{otherwise} \end{cases}
\label{ext2}
\end{equation} where $p_e$ is the error probability of the IDS.

In \cite{frigault} a dynamic BN is proposed to consider variations in time of the CVSS scores. Although the theoretical model in \cite{frigault} is correct, scores do not change rapidly over time, while additional complexity of the dynamic BN limits its application in real scenarios. In terms of computational complexity, it would perhaps be more reasonable to update the conditional probability distributions in the model we propose and recompute. To support this claim, the experimental results in Section 5 show the computational cost implications of augmenting the complexity of the BAG.

\section{Conclusions}
We have proposed in this paper a BN model for AGs and efficient exact inference techniques for their static and dynamic analysis. These techniques allow to assess the risk of the nodes in a network against cyber-attacks by calculating the probabilities that an attacker can compromise each node given the nodes that have been already compromised. These can help system administrators to respond to an attack or to take the corresponding countermeasures to an ongoing intrusion.

We have reviewed and proposed solutions to the shortcomings of existing BAG models in the literature, such as the implications of adding prior probabilities in the attacker's capabilities. In this sense, although some of our assumptions are still restrictive, we have proposed and sketched some direct extensions of the Bayesian model that we will consider in our future work.

We have also shown the limitations of previous state-of-the-art techniques to perform static and dynamic analysis of BAG, and the importance of using efficient algorithms to calculate the marginal probabilities. To support this, we have presented an extensive experimental evaluation with synthetic AGs to measure the time and memory required to calculate them. This stands in contrast with the related work where such analysis is missing. Our results show the advantages of the JT algorithm to perform static and dynamic analysis of AGs for graphs with hundreds of nodes, which in most cases would correspond to corporate networks of thousand of nodes. We have shown the important improvements of the JT algorithm in terms of time and memory with respect to the VE algorithm proposed in \cite{liu}. 


We have further shown the importance of modelling AGs taking into account the subnetwork structure of typical corporate networks. Network clustering enables the dynamic analysis with the JT algorithm to become tractable and scales linearly in the number of nodes. This allows to rapidly integrate new evidence in the analysis and enables administrators to respond to an ongoing attack.

Further directions include exploring more scalable inference techniques, extending the BAG model to make it less restrictive, and investigating more accurate means of estimating the probability of exploitation of vulnerabilities. Furthermore, beyond static and dynamic analysis, we can extend our model to other uses, such as prioritising forensic investigations and evidence collection as suggested in \cite{wijesekera}.


%

\section*{Acknowledgements}
This work has been supported by the UK government under EPSRC grant EP/L022729/1. The authors would like to thank British Telecom for their collaboration in this research and our colleagues in our research group for their contribution to this work through many useful discussions.


\ifCLASSOPTIONcaptionsoff
  \newpage
\fi



%
\bibliographystyle{IEEEbib}
\bibliography{biblio} 

%

\vspace{-1.2cm}

\begin{IEEEbiographynophoto}{Luis Mu\~noz-Gonz\'{a}lez}
received the PhD degree from University Carlos III of Madrid. He is currently a Research Associate at the Department of Computing at Imperial College London. His main research interests include Bayesian models, approximate inference, probabilistic graphical models, and machine learning applied to cyber security.
\end{IEEEbiographynophoto} 

\vspace{-1.2cm}

\begin{IEEEbiographynophoto}{Daniele Sgandurra}
holds a PhD in Computer Science from the University of Pisa. He is currently a Research Associate at the Department of Computing, Imperial College London. His main research fields include virtualization and cloud security, mobile security, threat modelling, social network security and privacy.
\end{IEEEbiographynophoto} 

\vspace{-1.2cm}

\begin{IEEEbiographynophoto}{Mart\'{i}n~Barr\`ere}
received his PhD degree in Computer Science from the University of Lorraine, France, in 2014. He is currently a Research Associate at the Department of Computing at Imperial College London. His main research interests include network and cloud security, autonomic computing, vulnerability management and digital forensics. 
\end{IEEEbiographynophoto} 

\vspace{-1.2cm}

\begin{IEEEbiographynophoto}{Emil C. Lupu}
leads the Academic Centre of Excellence in Cyber Security Research at Imperial College London. His work focuses on the engineering of resilient, adaptive and trustworthy systems across a broad range of systems ranging from IoT to Cloud environments. His prior work includes many contributions on policy-based network and systems management and security.
\end{IEEEbiographynophoto}

\end{document}